\documentclass[12pt]{article}
\usepackage[hmargin=2.5cm,vmargin=2.5cm]{geometry}
 \usepackage{graphicx}
 \usepackage{amsmath}
 \usepackage{amssymb}
  \usepackage{enumerate}
\usepackage{cite}
\usepackage{mathabx}
\usepackage{url}
\newcommand{\bey}{\begin{eqnarray}}
\newcommand{\eey}{\end{eqnarray}}
\usepackage{etoolbox}
\usepackage{tabularx}
\usepackage{lipsum}

\begin{document}
\title{Relativistic Quantum Information from Unequal-Time QFT Correlation Functions}
\author {Charis Anastopoulos  \thanks{anastop@upatras.gr} \;
and   Konstantina Savvidou  \thanks{ksavvidou@upatras.gr}
 \\ \\
\noindent   {\small Department of Physics, University of Patras, 26500 Greece}
 }

\maketitle
\begin{abstract}
This paper continues on the program of developing a relativistic quantum information theory in terms of unequal-time correlation functions in quantum field theory (QFT) [C. Anastopoulos, B. L. Hu, and K. Savvidou, Ann. Phys. 450,  169239 (2023)]. 
Here, we focus on the definition of quantum resources from the irreducibly quantum behavior contained in the correlation functions of a QFT. We explain how set-ups with $N$ particle detectors probe the information in the high order field correlation functions. Our main object is the associated hierarchy of probability densities of $N$-detector events. We show that classical probabilistic hierarchies are subject to two conditions: Kolmogorov additivity and measurement independence. QFT violates those conditions, and the degree of violation enables us to define novel quantum resources. We give specific examples in set-ups where the main observables are the times of particle detection events. The new resources capture instances of irreducibly
 quantum behavior that differs from the quantum behavior encapsulated in Bell inequalities. An interesting byproduct of our analysis is a relativistic state reduction rule for particles detected through scattering.

\end{abstract}

\section{Introduction}
 
Understanding the relation of quantum information theory (QIT) to relativity is a major and multi-faceted task.  
 On one hand, it involves the application of
quantum information  concepts to fundamental problems of relativistic physics, as they appear in  diverse
research fields, including particle physics, cosmology, quantum gravity, and quantum optics. On the other
hand, it requires the incorporation of the fundamental principles of relativity, namely, causality and
covariance, into the foundations of QIT. This will create novel opportunities for both theory and  technological applications.

To achieve those ends, the grounding of quantum information concepts on
 quantum field theory (QFT) is crucial. QFT   is the only known theory that combines quantum theory and special relativistic dynamics consistently, and
in agreement with experiments. Its domain of validity extends along the full range of physics,  from quarks to condensed matter, to black holes, and  the early universe. So far, the mainstream QIT
paradigms have been developed in the context of non-relativistic quantum mechanics, which is a small corner of
full QFT. 

Certainly, a transfer of non-relativistic concepts to relativistic setups is possible, and, indeed,
fruitful. For example, some aspects of entanglement are commonly discussed in   QFT, especially in relation to black hole information and holography---see, for example, \cite{CaCa06, Witten}. However, entanglement in QFT does not have the informational interpretation that it has in non-relativistic quantum information theory. In the latter context, entanglement is identified as a quantum resource associated to the  Local Operation and Classical Communication (LOCC) protocol \cite{LOCC}. There is no relativistic analogue for LOCC, mainly due to our limited knowledge about admissible quantum operations in QFT, 
 especially in relation to relativistic locality and causality. Hence, a first-principles informational interpretation of entanglement in QFT, in terms of information obtained from measurements is missing. Furthermore, relativistic covariance suggests the need of a unified treatment of spacelike quantum correlations (as in Bell non-separability \cite{Bell}) and temporal ones (as in the violation of the Leggett-Garg inequalities \cite{LeGa}).

\subsection{The Quantum Temporal Probabilities approach}

This paper is part of a larger program to develop a consistent relativistic QIT from first principles \cite{AnSav22, AHS23}. This program emphasizes the need of  a consistent and practicable measurement theory for relativistic QFT. This is a long-standing problem in the foundations of QFT. A major part of the problem is the lack of a consistent relativistic rule for state update after measurements. A naive transfer of the rule for quantum state reduction used in non-relativistic quantum theory to a relativistic setup consists with  relativistic causality ---see \cite{PeTe, AnSav22, AHS23, MD} and references therein.

 Modern QFT has sidestepped such problems by focusing on  scattering experiments that can be treated via S-matrix theory---see Refs.  \cite{Blum,MD}.  The latter treats scattering as a process with a single measurement event in the asymptotic future, so the state-update rule is not used. However,  
a state-update rule is needed to construct joint probabilities for multiple measurement events, and no probabilistic theory can work in its absence. It is required in the analysis of Bell-type experiments, experiments involving temporal quantum correlations,  and for post-selected measurements. 

In particular, quantum optics requires joint probabilities, to describe  photon bunching and anti-bunching. To this end, photo-detection theories have been developed, the most prominent of which is that of Glauber \cite{Glauber1, Glauber2}. 
Glauber's theory was arguably the first QFT measurement theory, and it proved remarkably successful. However, its scope is limited.  It works only for photons, and it involves approximations that may not be consistent with relativistic causality.  

Early attempts to describe QFT measurements include the formal  treatment of Hellwig and Kraus \cite{HK},  the development of detector models by Unruh and DeWitt \cite{Unruh76, Dewitt}, and Sorkin's important argument about the conflict between ideal measurements and QFT causality \cite{Sorkin}. 
Recent years have witnessed renewed interest on the topic; see, for example, Refs. \cite{QTP1,  QTPdet, OkOz, FeVe, Bostelmanetal, FJR22, GGM22, Perche, PTM, Bednorz, PRA24}; see also Refs. \cite{AHS23, MD, FeVe2} which review aspects of those developments.
In this work, we describe QFT measurements using the Quantum Temporal Probabilities method (QTP) \cite{QTP1, QTP2, QTP3}. The original motivation of QTP was to provide a general framework for  temporally extended quantum observables \cite{AnSav06},  hence  the name. 
The key idea  in QTP is to distinguish between the time parameter of Schr\"odinger's equation from the time variable associated to  particle detection \cite{Sav99, Sav10}. The latter is then treated as  a macroscopic  variable associated to the detector degrees of freedom.

A major difference of QTP from other approaches to QFT measurements is that it describes measurements within the decoherent histories approach to quantum mechanics \cite{Gri, Omn1, Omn2, GeHa1, GeHa2, hartlelo}, rather than trying to connect with von Neumann's measurement theory \cite{vN, Busch}. This means that QTP  focusses on the emergent classicality in the measurement apparatus. Observables correspond to coarse-grained histories of the apparatus' degrees of freedom that satisfy a decoherence condition.

 QTP emphasizes that every measurement event is localized in spacetime, and that the time of its occurrence is a random variable. This contrasts the von Neumann description  of quantum  measurements, in which the time of a measurement event is {\em a priori} fixed. The QTP treatment is closer to   experimental practice. Actual particle detectors are fixed in space and the timing of their recordings varies probabilistically.  In QTP, the random variables for each detection event are of the form $(x, q)$, where $x$ is the spacetime point of detection, and $q$ refers collectively to all other observables (particle momentum, angular momentum, and so on). Hence, QTP provides probabilities densities of the form
\bey
P_n(x_1, q_1; x_2, q_2; \ldots ; x_n, q_n), \label{pnnn}
\eey
 associated to set-ups with $n$ separate detectors. 
 
 The probability densities (\ref{pnnn}) 
 are linear functionals of unequal time correlation of the fields with $2n$ entries. 
  This is particularly important, because such correlation functions are a staple of QFT, and there exists a powerful arsenal of techniques for their calculation.  The specific correlation functions relevant to QTP are not the usual ones of S-matrix theory (in-out formalism), but they appear in the Closed-Time-Path (CTP) (Schwinger-Keldysh or `in-in') formalism \cite{ctp1, ctp2, ctp3, ctp4, ctp5}. Unlike $S$-matrix theory, the CTP formalism
allows for the evaluation of real-time---rather than asymptotic---probabilities for the quantum fields. It has found   many applications in  nuclear-particle physics, early Universe cosmology, and condensed matter physics \cite{CH08, Berges, coma1, coma2}. The connection between the two formalisms 
allows us to combine the tools  of quantum measurement theory and QFT, that is, the language of    Positive-Operator-Valued measures (POVMs) and effects, with the manifestly covariant language of  path integrals.

\subsection{Quantum information from the correlation hierarchy}

A crucial fact about QFTs is that all information is encoded into hierarchies of unequal-time correlation functions.  For example, Wightman's reconstruction theorem \cite{Wightman1, WiSt} allows one to construct the full QFT and the ground state from the properties of a theory's Wightman functions, that is, vacuum expectation values of products of field operators. The Lehmann-Symanzik-Zimmermann construction enables the calculation of all elements of the S-matrix from the hierarchy of time-ordered correlation functions \cite{LSZ}. In the general case, all real-time properties of a QFT can be constructed from the CTP hierarchy of unequal-time field correlation functions \cite{cddn, DCH}
\bey
G^{m, n}(x_1, \ldots, x_n; x_1', \ldots x'_m) = Tr \left[{\cal T}^*[\hat{\phi}(x_n) \ldots \hat{\phi}(x_1)]\hat{\rho}_0{\cal T}[\hat{\phi}(x_1) \ldots \hat{\phi}(x_n)] \right],
\eey
where $\hat{\phi}$ is an arbitrary field, $\hat{\rho}_0$ is the initial state of the field, ${\cal T}$ stands for time-ordering and ${\cal T}^*$ for reverse time ordering. 

\begin{figure}
 \includegraphics[width=12.6cm]{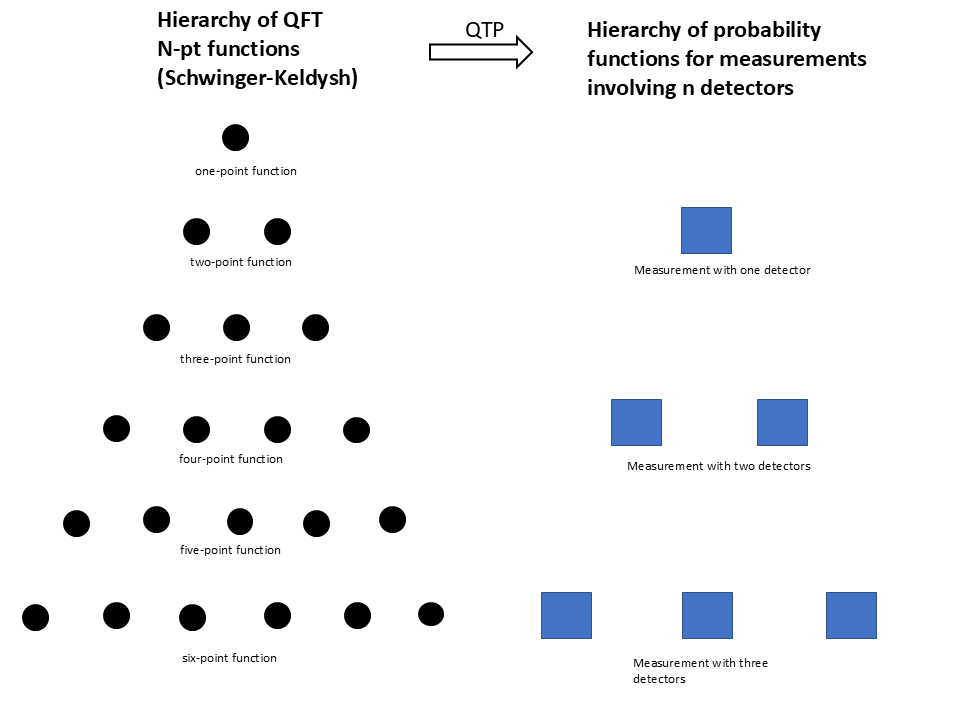}\label{hierar}
     \caption{The hierarchy of correlation functions in relation to the hierarchy of probabilities for multi-detector measurements.}
\end{figure}

This means that the hierarchy of the unequal-time correlation functions contains all information about the QFT, including  Bell-type and dynamical quantum correlations. The starting point of our analysis is that this information can be extracted through QTP-type measurements. For each type of detector that records events of the form $(x, q)$, there exists a hierarchy of probability densities 
 (\ref{pnnn}). Each member of the hierarchy corresponds to a measurement set-up with $n$ detectors of this type, and it extracts information corresponding to the correlation function $G^{n,n}$ of the CTP hierarchy---see Fig. \ref{hierar}.  
 
In this paper, we identify the irreducibly quantum characteristics in QTP probability hierarchies, and, consequently, of the associated CTP field correlation functions. The key observation is that classical probabilistic hierarchies are subject to constraints, due to compatibility conditions between the different levels of the hierarchy. We show that these constraints are violated in quantum theory. Then, the size of this violation, as quantified by a norm, defines a quantum resource associated to the probabilistic hierarchy. 

We identify two types of classicality conditions in probabilistic hierarchies, {\em Kolmogorov additivity} and {\em measurement independence}.

\medskip

\noindent 1.  A probabilistic hierarchy can be obtained from a single stochastic probability measure, if it satisfies  Kolmogorov additivity. This is a compatibility condition between any level of the hierarchy and its lower ones. Quantum theory can violate Kolmogorov additivity, a feature that has also been identified in non-relativistic set-ups \cite{Ana06, BrKo, ClKo}, and it is closely related to the violation of the Leggett-Garg inequalities \cite{Halli1, Halli2}. 

    \medskip
    
\noindent 2. A hierarchy that satisfies Kolmogorov additivity is described by a stochastic process, but this process is not necessarily local. The appropriate locality condition is measurement independence \cite{MI1, MI2, MI3}, which roughly asserts that the physical quantity that is being measured in a detector is defined independently of any other quantity that is being measured in the same set-up. In non-relativistic physics, measurement independence is a prerequisite for the derivation of Bell's inequality.

\medskip

The two conditions above constrain probability densities from different levels in the probabilistic hierarchy. This contrasts Bell-type inequalities \cite{Bell} that--in the present context--require the comparison of probability densities from {\em different} probabilistic hierarchies. For example, the derivation of the Clauser-Horne-Shimony-Holt (CHSH) inequality \cite{CHSH} considers four different pairs of detectors, and each pair would have to be embedded at the level $n = 2$ of a different probabilistic hierarchy.

In this paper, we analyze the implications of the classicality constraints above, and then we define the quantum resources defined by their violation. 
 For concreteness, we evaluate the latter for the simplest probabilistic hierarchy. This consists of probability densities of the form (\ref{pnnn}), in which the only recorded observable is the spacetime point of the detection event. Hence, the hierarchy consists of probability densities of the form $p_n(x_1, x_2, \ldots, x_n)$ for $n$ detection events at spacetime points $x_1, x_2, \ldots, x_n$. Such probability densities enable us to undertake an analysis of Einstein causality in QFT measurements. The analysis simplifies significantly when considering a set-up of small detectors at fixed spatial locations. Them, the only random variables are the detection times $t_1, t_2, \ldots, t_n$ in each apparatus.

We find that the violation of both constraints is generic, and, in principle, measurable. Kolmogorov additivity is violated in set-ups where particles are detected sequentially by two detectors, and it is due to the change of the quantum state after the first detection. Measurement independence is violated in multi-partite systems. In the cases analyzed here, it is related to  entanglement between recorded particles. 

\medskip 

The plan of this paper is the following. In Sec. 2, we summarize the main points of the QTP method for measurements in QFT. In Sec. 3, we analyse the classicality conditions of probabilistic hierarchies and we define the corresponding quantum resources. In Sec. 4, we focus on the violation of measurement independence in a bipartite system. In Sec. 5, we analyse the violation of Kolmogorov additivity. In Sec. 6, we summarize and discuss our results.

\section{Background}
In this section, we present the main points of the QTP formalism, as presented in Ref. \cite{AHS23}, and we setup the notation for the rest of this paper.

Consider a general quantum field theory defined by a Hilbert space ${\cal F}$. The quantum field interacts with $n$ detectors, each localized in different spatial region, and described by the Hilbert space ${\cal H}_i$, for $i = 1, 2, \ldots, n$. Each detector records the time and location of a particle-detection event, and also the value of 
additional observables (e.g., four momentum, angular momentum), collectively labeled by $q$. The state of motion of the detectors may be arbitrary. 

The quantum field interacts with the $i$-th detector through an interaction term of the form
\bey
\hat{V}_i = \int d^4 x \hat{C}^{(i)}_a(x) \hat{J}_{(i)}^a(x),
\eey
where $\hat{C}_{a}^{(i)}(x_i)$ is a field composite operator that defines the channel of interaction between field and detector and here $\hat{J}_{(i)}^a(x)$ is a (Heisenberg-picture) current on the detector  Hilbert space ${\cal H}_i$. The index $a$ is composite, and it may include spacetime, spinor, and internal indices. Raising and lowering the indices occurs, so that expressions of the form $K_aL^a$ are spacetime scalars.

Some examples of relevant composite operators include 
\begin{itemize}
\item  dipole coupling  $\hat{C}(x) = \hat{\phi}(x)$ for a scalar field $\hat{\phi}(x)$,
\item coupling for detection through scattering $\hat{C}(x) = :\hat{\phi}^2(x):$ for a scalar field $\hat{\phi}(x)$,
\item scalar and vector couplings $\hat{C}(x) = :\hat{\bar{\psi}}(x)\hat{\psi}(x):$ and $\hat{C}^{\mu}(x) = :\hat{\bar{\psi}}(x)\gamma^{\mu}\hat{\psi}(x):$, respectively,  for a Dirac field $\hat{\psi}(x)$.
\end{itemize}

The QTP analysis leads to an unnormalized  probability density for the $n$ measurement events,
\bey
P_n(x_1,q_1; x_2, q_2; \ldots; x_n, q_n) = \int d^4 y_1 \ldots d^4 y_n     R_{(1)}^{a_1b_1}(y_1, q_1) \ldots R_{(n)}^{a_nb_n}(y_n, q_n)
\nonumber \\
\times G_{a_1 \ldots a_n, b_1\ldots b_n}(x_1 - \frac{1}{2}y_1, \ldots, x_n - \frac{1}{2}y_n; x_1 + \frac{1}{2}y_1, \ldots, x_n + \frac{1}{2}y_n), \label{probden4}
\eey
where $x_i$ are spacetime coordinates of a detection event. The probability density is a linear functional of the $2n$-unequal-time correlation function  
and
\bey
G_{a_1 \ldots a_n, b_1\ldots b_n}(x_1, \ldots, x_n; x_1', \ldots, x_n') = \langle \psi| {\cal T}^*[\hat{C}_{b_1}^{(1)}(x'_1) \ldots \hat{C}_{b_n}^{(n)}(x'_n) ]
\nonumber \\
\times {\cal T} [\hat{C}_{a_n}^{(n)}(x_n) \ldots \hat{C}_{a_1}^{(1)}(x_1)]|\psi\rangle \label{correl}
\eey
 of the field composite operators $\hat{C}_{a_i}^{(1)}(x_i)$.

The quantities $R^{ab}_{(i)}(x, q)$ in Eq. (\ref{probden4})
 are the {\em detector kernels} of each measurement. They contain all information pertaining to the detector, including its internal dynamics and its state of motion. The expressions for the detector kernel simplify under the following two assumptions.
 
 \begin{enumerate}
 \item A detector follows an inertial trajectory in Minkowski spacetime.
 \item A detector is initially prepared on a state $|\Omega\rangle$ that is  {\em approximately translation invariant}. Heuristically, this means  that the apparatus is prepared in an state that is homogeneous at the length scales relevant to position sampling and approximately static at the time scales that relevant to time sampling. 
 \end{enumerate}
 Then, the detector kernel takes the form 
 \bey
R^{ab}(x, q) = \langle a, q|e^{i \hat{p} \cdot (x - x_0)}|b, q\rangle,
\eey
 where $x_0$ is a reference point in the world-tube of the detector, and $|a, q\rangle = \sqrt{\hat{\Pi}(q)} \hat{J}(x_0)|\Omega\rangle$. Here, 
 $\hat{\Pi}(q)$ is a set of positive operators on the Hilbert space of a single detector, that is correlated to the values of the observable $q$. It satisfies $\sum_q \hat{\Pi}(q) = \hat{I} - |\Omega\rangle \langle \Omega|$.
 Some explicit models for detector kernels are described in the Appendix B.

The Fourier transform of the detector kernel
\bey
\tilde{R}^{ab}(\xi, q) = \int d^4x e^{-i \xi\cdot x} R^{ab}(x, q),
\eey
is given by
\bey
\tilde{R}^{ab}(\xi, q) = (2\pi)^4 e^{i \xi\cdot x_0} \langle a, q|\hat{E}_{\xi}|b, q\rangle, \label{tildr}
\eey
where $\hat{E}_{\xi} = \delta^4(\hat{p} - \xi)$ is the projector onto the subspace with four-momentum $\xi^{\mu}$.
The momentum four vector associated to the detector is timelike and the associated energy $p^0$ is positive. This means that $\tilde{R}^{ab}(p, q) = 0$ for spacelike $p$, or if $p^0 < 0$. Eq. (\ref{tildr}) also implies that $\tilde{R}^{ab}(\xi, q) \geq 0 $, since $\langle a, q|\hat{E}_{\xi}|a, q\rangle \geq 0$.

It is convenient to express the probability densities  using an abstract index notation. We use small Greek indices $\alpha, \beta, \gamma \ldots$ for the pairs  $(x, a)$ where $x$ is a spacetime point and $a$ the internal index for the composite operators $\hat{C}_a$. All indices  in a time-ordered product are upper, and all indices in an anti-time-ordered product are lower. Hence, we write the correlation functions (\ref{correl}) as
\bey
G^{\alpha_1 \alpha_2 \ldots \alpha_n}{}_{\beta_1 \beta_2 \ldots \beta_n} \nonumber
\eey

We denote by $z$ the  pairs $(x, q)$, where $x$ is a spacetime point and $q$ any other recorded observable. In this shorthand, we include in $z$ also the event $\emptyset$ of no detection. If we denote by $\Gamma$ the set of possible values for $q$, then $z$ takes values on the set $Z:= M \times \Gamma \cup \{\emptyset\}$.
We will write the kernel
\bey
\delta[x - \frac{1}{2}(y+y')]R^{ab}(y - y', q) \label{raabb}
\eey
as $R_\alpha{}^\beta(z)$ where $\alpha$ stands for $(y, a)$, $\beta$ for $(y', b)$ and $z$ for $(x, q)$. The matrices $R_\alpha{}^\beta(z)$ are self-adjoint by construction: $R^{\alpha}{}_{\beta} = (R_{\alpha}{}^{\beta})^*$.

Then, the probability formula (\ref{probden4}) becomes
\bey
P_n(z_1, z_2, \ldots, z_n)  = G^{\alpha_1 \alpha_2 \ldots \alpha_n}{}_{\beta_1 \beta_2 \ldots \beta_n} \;\; {}^{(1)} R_{\alpha_1}{}^{\beta_1}(z_1) \; {}^{(2)} R_{\alpha_2}{}^{\beta_2}(z_2) \ldots \; {}^{(n)} R_{\alpha_n}{}^{\beta_n}(z_n). \label{probdenr}
\eey
In this notation, each measurement event is associated to a pair of indices in the correlation function, an upper one $\alpha_i$ and a lower one $\beta_i$. 

In some cases, it will be useful to further condense the notation, and employ DeWitt notation, in which each pair  $(\alpha_i, \beta_i)$ corresponds to a single capital index $A_i$. Hence, we will write the 2n-point correlation function as $G_{A_1,\ldots, A_n}$ and the detector kernels as $R^A(z)$. Then, the probability formula (\ref{probdenr}) reads
\bey
P_n(z_1, z_2, \ldots, z_n)  = G_{A_1 A_2\ldots A_n}   R_1^{A_1}(z_1)  R_2^{A_2}(z_2) \ldots R_n^{A_n}(z_n) \label{probdenr2}
\eey
 Each vector $v_A$ corresponds to a matrix $v_{\alpha}{}^{\beta}$, and we can define the Hilbert-Schmidt inner product  
\bey
(v, w) = Tr(vw^{\dagger}).
\eey
Hence, the  space ${\cal V}$ of vectors $v_A$ with finite norm $||v|| = \sqrt{v_A \bar{v}^A}$  defines a Hilbert space. The   inner product allows us to raise and lower the capital indices consistently. 
In particular, we will denote the Riesz-dual of a vector $v_A$ by $\bar{v}^A$.   We will also define the norm of any ``tensor`` $G_{A_1, \ldots, A_n}$ as $||G|| := \sqrt{G_{A_1, \ldots, A_n} \bar{G}^{A_1, \ldots, A_n}}$.

\section{Measures of non-classicality}
In this section, we analyze the structure of the hierarchy of QTP probability functions, and we identify behavior with no classical analogue. We also define informational quantities that quantify the degree if irreducible non-classicality.

 \subsection{Non-additivity in the hierarchy of probability densities}

Suppose that we have a specific rule for selecting the type of apparatuses that appear in the $n$-event measurement described by the probability distribution  (\ref{probden4}), for all $n = 1, 2, \ldots$. The simplest such rule is to assume that all apparatuses are identical, modulo a spacetime translation of their worldtubes, so that the associated detector kernels differ only by a phase term $e^{i\xi\cdot a}$ in the Fourier space. We can absorb such phase differences in a redefinition of the observables $z_i$, so that $R^A_i (z) = R^A(z)$ for all $i$.

Given such a rule, Eq. (\ref{probdenr}) assigns to each initial state $|\psi\rangle$ of the field a hierarchy of joint  probability distributions  $P_n(z_1, z_2, \ldots, z_n)$, where $n = 1, 2, \ldots, $.
We can define a moment-generating functional for all probability densities
 (\ref{probdenr}), in terms of sources $j(z)$,
\bey
Z_{QTP}[j] = \sum_{n=0}^{\infty} \sum_{z_1, z_2, \ldots, z_n} \frac{i^n}{n!}  P_n(z_1, z_2, \ldots, z_n) j(z_1) \ldots j(z_n), \label{zqtp}
\eey
where we took  $p_0 = 1$.

 In classical probability theory, a hierarchy of correlation functions defines a  stochastic process, if it satisfies the Kolmogorov additivity condition,
\bey
P_{n-1}(z_1, \ldots,z_{i-1}, z_{i+1},\ldots, z_{n}) = \int dz_i  P_n(z_1, \ldots,z_{i-1}, z_i, z_{i+1},\ldots,  z_n) \label{Kolmadd}
\eey
for all $i = 1, 2, \ldots n$ and $n = 1, 2, \ldots$. 

Quantum probability distributions for sequential measurements do not satisfy this condition \cite{Ana06}. Typical probabilities for non-relativistic measurements are of the form   
\bey
P_2(z_1, z_2) = \langle \psi|\hat{P}_{z_1} \hat{Q}_{z_2}\hat{P}_{z_1}|\psi\rangle, \label{nonrel}
\eey
for $n=2$, 
where $\hat{P}_{z_1}$ and $\hat{Q}_{z_2}$ are the spectral projectors associated to the first and the second measurement, respectively. If the projectors $\hat{P}_z$ and $\hat{Q}_{z'}$ do not commute, then the Kolmogorov condition is typically violated. QTP probabilities have a more complicated structure, but they follow a similar pattern. The violation of the Kolmogorov condition is related the  violation of the 
 Leggett-Garg inequalities for macrorealism  \cite{LeGa} that also refer to the behavior of quantum multi-time probabilities.

The violation of Eq. (\ref{Kolmadd}) is a genuine signature  of quantum dynamics; it cannot be reproduced by any classical stochastic processes. 
In contrast, if measurements on a quantum field approximately satisfy Eq. (\ref{Kolmadd}), then the measurement outcomes can be  simulated by a stochastic process with $n$-time probabilities given by the probability distributions (\ref{Kolmadd}). Then, the generating functional $Z_{QTP}$ is the  functional Fourier transform of a classical stochastic probability measure.

Hence, Eq. (\ref{Kolmadd}) provides unambiguous classicality criterion. The divergence of a probabilistic hierarchy  from  Eq. (\ref{Kolmadd})  is a measure of the hierarchy's ``quantumness", and it can be interpreted as a quantum resource.  To quantify this resource, we define the marginals of the probability density $P_n$,
\bey
\tilde{P}_{n, i}(z_1, z_2, \ldots, z_n) = \int d\zeta  P_{n+1}(z_1, \ldots,z_{i}, \zeta, z_{i+1},\ldots,  z_n).
\eey
We denote by $w_{n, i}$ the statistical distance between the probability distributions  $\tilde{P}_{n,i}$ and $P_n$,
\bey
w_{n,i} = \delta(\tilde{P}_{n,i}, P_n) = \frac{1}{2} \int d^nz |\tilde{P}_{n,i}(z_1, \ldots, z_n)  - P_{n}(z_1, \ldots, z_n) |.
\eey
The smallest value for $w_{n, i}$ is $0$, obtained when $\tilde{P}_{n,i} = P_n$. The largest possible value for $w_{n_i}$ is $1$, obtained when  $\tilde{P}_{n,i}$ and $P_n$ have disjoint supports. We define the {\em non-additivity} of the $n$-th level of the probabilistic hierarchy as  
\bey
w_n = \frac{1}{n} \sum_{i=1}^n w_{n, i},
\eey
with values in $[0, 1]$.

In many cases, we are only interested in the lower levels of the hierarchy. For example, if we use the hierarchy in order to describe the non-equilibrium thermodynamics of quantum fields, Boltzmann's equation arises at the level of $n = 1$, so $w_2$ suffices as a measure of non-additivity. To quantify non-additivity in the whole hierarchy, it suffices to specify a finite norm for the sequence    $\{w_n\}$. Since the sequence is bounded, it is convenient to use the supremum norm, so we define the non-additivity of the hierarchy
\bey
W = \sup_n w_n.
\eey
which is guaranteed to be finite. 

The non-additivity $W$ is defined to generic probabilistic hierarchies, and not only to the ones expressed by the QTP formula. In general, it is possible to exploit the QTP  formula and define non-additivity measures that are expressed in terms of QFT correlation functions. We show how this works, by focussing at the 
  level $n=2$ of the hierarchy. The probability densities  $P_1$ and $P_2$ are given by
\bey
P_1(z) &=& C_1 G_{A} R^A(z) \\
P_2(z_1, z_2) &=& C_2 G_{AB} R^A(z_1) R^B(z_2),
\eey
where $C_1$ and $C_2$ are constants that implement the normalization conditions $\int dz P_1(z) = \int dz_1 dz_2 P_2(z_1, z_2) = 1$. Here, we chose $R_1^A = R_2^A$

Consider now the two marginals of $P_2$,
 \bey
 P_{21}(z) = \int dz_2 P_2(z, z_2) = C_2 G_{AB} R^A(z) D^B, \\
   P_{22}(z) = \int dz_1 P_2(z, z_1) = C_2 G_{AB}D^AR^B(z) ,
 \eey
where $D^A= \int dz R^A(z)$ is a kernel that accounts for the total number of detection events on the $i$-th detector. For $R^A(z)$ of the form (\ref{raabb}), $D^A$ corresponds to the kernel $\int dq R^{ab}(y-y', q)$. It describes detectors that records only the spacetime coordinate of an event.

Kolmogorov additivity holds only if
\bey
C_2 G_{AB} D^B = C_2 G_{BA} D^B = C_1 G_A \label{kolgad}
\eey
%This condition obviously holds if the four-point correlation factorizes into a pair of two point functions, i.e., if $G_{AB} = G_AG_B$, but it may also hold for non-factorized correlation functions. For example, suppose that the initial state $\hat{\rho}_0$ of the field can be expressed as  $\sum_a \lambda_a \hat{\rho}_0^{(a)}$, where   $\sum_a \lambda_a = 1$, and $\hat{\rho}_0^{(a)}$ are density matrices. Note that $\lambda_a$ are not restricted to be positive, so $\hat{\rho}_0$ is not necessarily a convex combination of the density matrices $\hat{\rho}_0^{(a)}$ .
%Suppose that the correlation functions associated to the density matrices $\hat{\rho}^{(a)}$, satisfy  $G_{AB}^{(a)} =  G_A^{(a)}G_B^{(A)} $.  Then, the correlation functions $G_{AB} = \sum_a \lambda_a  G_{AB}^{(a)} $ and $G_A = \sum_a \lambda_a G_A^{(a)}$ also satisfy Eq. (\ref{kolgad}). This means that the quantum behavior manifested in the violation of Eq. (\ref{kolgad}) is not due to any properties of the quantum state, but rather to the type of measurement that is performed on the system.

%To see this, note that the quantum probability density for two local measurements in a bipartite system is given by
%\bey
%P(z_1, z_2) = \langle \psi|\hat{\Pi}(z_1) \otimes \hat{\Pi}(z_2)|\psi\rangle, \label{factorme}
%\eey
%where $\hat{\Pi}(z)$ is  a positive operator valued measure (POVM) associated to one local measurement. It is straightforward to show that this class of measurements satisfies Eq. (\ref{kolgad}) for all choices of the initial state. 

Eq. (\ref{kolgad}) is equivalent to the statement that the two vectors $G_{AB} D^B$ and  $G_{BA} D^B$ coincide, and they are parallel to $G_A$. This means that we can quantify the violation of the Kolmogorov condition by two numbers,

\begin{itemize}
\item the norm of the vector $G_{[AB]}D^B$, and 
\item the angle $\theta$ between the vectors $G_{BA} D^B$ and $G_A$, defined by
\bey
\cos \theta = \frac{G_{BA}D^B \bar{G}^A}{||G_A|| \, ||G_{BA} D^B||}.
\eey 
\end{itemize}

In many cases, $G_{AB} D^B$ is indeed parallel to $G_A$, because  tracing out the last measurement does not affect the previous outcomes (unless the probabilities are post-selected). Then, all information about non-additivity is contained into the angle $\theta$. 

%The failure of Kolomogorov additivity is due to the fact that the measurements are not fundamentally separable, even if the apparatuses are located far from each other.

\subsection{Violation of measurement independence}

If Kolmogorov additivity is satisfied, then the hierarchy of probability functions is described by a stochastic process. This means that the probabilities for a setup involving $n$ detectors remain unchanged if we add yet another detector. In non-relativistic quantum theory, this means that the observables that are being measured always commute with each other. 

Certainly, the existence of a stochastic process that simulates measurement outcomes does not guarantee classicality. Classicality requires additional conditions that enforce the separability of measurements. Of particular importance is the notion of {\em measurement independence}. Measurement independence (MI) denotes the
statistical independence of any parameter that affects the selection of
measurement procedures from physical variables that influence the measurement
outcomes. MI enters explicitly the derivation of Bell inequalities\footnote{MI is called different names by different authors. Here, we use the terminology of \cite{MI2}---see also \cite{Ana23}.} It
 is significantly weaker than Bell locality (BL), namely, the assumption that the result of a measurement on one system be unaffected by operations on any  distant (spacelike separated) system. MI implies BL, but the converse does not hold. 
 
 In the present context, we implement MI as follows. Suppose that the stochastic process is described by the random variables $\xi$; typically, $\xi$ are elements of a space $\Pi$ of paths over a single-time sample space. Observables are functions on $\Pi$. Suppose that we measure an observable   $Z$ with values $z$; each measurement outcome is represented by a subset $C_z$ of $\Pi$. Let us denote by $F_z$, the characteristic function of $C_z$. By definition, $F_z^2 = F_z$, and $Z = \int dz z F_z$. For a statistical ensemble of measurements described by a probability  density  $\rho(\xi)$, the probability density 
\bey
P_1(z) = \int d\xi \rho(\xi) F_z(\xi) \label{FzP}
\eey
describes measurements of the observable $Z$.
For   unsharp measurements,  the condition $F_z^2 = F_z$ can be relaxed; it suffices that $F_z$ be positive.

Suppose that we make $n$ independent measurements of observables $Z_1, Z_2, \ldots, Z_n$. MI implies that we can assign a different function $F^{(i)}_{z_i}(\xi)$ to each observable. Then, the joint probability density $P_n(z_1, z_2, \ldots, z_n)$ takes the form
\bey
P_n(z_1, z_2, \ldots, z_n) = \int d \xi \rho(\xi) F^{(1)}_{z_1}(\xi) F^{(2)}_{z_2}(\xi) \ldots F^{(n)}_{z_n}(\xi).
\eey
 Probability densities of this form are severely constrained.
  To see this, assume that it is possible to have a set-up where $z_2 \rightarrow z_1$. Then, by Jensen's inequality,
  \bey
  P_n(z_1, z_1, \ldots, z_n) \geq P_{n-1}(z_1, z_3, \ldots, z_n)^2 \label{pepe}
  \eey
  For $n=2$, Eq. (\ref{pepe}) becomes
  \bey
  P_2(z, z) \geq P_1(z)^2. \label{p221}
  \eey

Setting $A(\xi) = \sqrt{\rho(\xi)} F^{(1)}_{z_1}(\xi) \ldots F^{(m)}_{z_m}(\xi)$, $B(\xi) = \sqrt{\rho(\xi)} F^{(m+1)}_{z_1}(\xi) \ldots F^{(n)}_{z_m}(\xi)  $ 
for $m < n$, we write  $P_n = \int d\xi A(\xi) B(\xi)$.Then, the 
     Cauchy-Schwartz inequality yields 
\bey
P_n(z_1, z_2, \ldots, z_n) \leq \sqrt{P_{2m}(z_1, z_1, \ldots, z_m, z_m) P_{2n-2m}(z_{m+1}, z_{m+1}\ldots, z_n, z_n)}. \label{quant2}
\eey
In particular, for $n = 2$, 
\bey
P_2(z_1, z_2) \leq \sqrt{P_2(z_1, z_1)P_2(z_2, z_2)}. \label{CauSch}
\eey
The inequalities (\ref{quant2})  cut across different levels of the probabilistic hierarchy in a QFT. 
 They are loosely analogous to the violations of classicality conditions of the quantum electromagnetic field, in phenomena such as photon anti-bunching.

%If we make the more stringent assumption that $F_z^2 = F_z$, Eq. (\ref{pepe}) becomes $P_n(z_1, z_1, \ldots, z_n) = P_{n-1}(z_1, z_3, \ldots, z_n)$. Eq. (\ref{quant2}) simplifies to
%\bey
%P_n(z_1, z_2, \ldots, z_n) \leq \sqrt{P_m(z_1, \ldots, z_m) P_{n-m}(z_{m+1}, \ldots, z_n)}. \label{quant2b}
%\eey

 The simplest way to quantify quantum behavior is through the norm of the inequality violating terms. To this end, define 
\bey
Q^{(1)}_n :&=& \int dz_1dz_3 \ldots dz_n \max\left\{0, P_{n-1}(z_1, z_3, \ldots, z_n)^2 - P_n(z_1, z_2, \ldots, z_n) \right\} \; \; \;\label{qn1}\\
Q^{(2)}_{n,m} :&=& \int dz_1 \ldots dz_n \max\left\{0,  P_n(z_1, z_2, \ldots, z_n) \right. \nonumber \\ &-& \left.  \sqrt{P_{2m}(z_1, z_1, \ldots, z_m, z_m) P_{n-m}(z_{m+1}, z_{m+1}\ldots, z_n, z_n)}\right\}. \label{qnm2}
\eey
We can average $Q^{(2)}_{n,m}$ over all values of $m$ of fixed $n$, so that the number
\bey
Q^{(2)}_n = \frac{1}{m} \sum_{m=1}^n Q^{(2)}_{n,m}
\eey
quantifies the violation of measurement independence at the $n$-th level of the hierarchy. Then, the supremum norms over the sequences $\{Q^{(1)}_n\}$ and $\{Q^{(2)}_n\}$  serve as   classicality measures for the whole hierarchy. 
 
The across-levels inequalities (\ref{pepe}, \ref{quant2}) are fundamentally different from Bell inequalities. The latter involve probabilities at the same level, but from different probabilistic hierarchies. For example, the  CHSH  inequality works at the level of $n = 2$, and it connects four different probability densities $P_2(z_1, z_2)$, each corresponding to a different set-up of the measurement apparatuses. The possible scenarios for the probabilistic hierarchies in QFT are sketched in Table \ref{TQ}.
 
%\begin{figure}
% \includegraphics[width=10.6cm]{tableQ}\label{TQ}
%     \caption{Alternatives about the quantum behavior of a hierarchy of probability distributions in QFT.}
%\end{figure}

\begin{table}\label{TQ}
 \begin{tabularx}{\textwidth}{|X|X|} 
   \hline
  
  {\bf Probabilities in a hierarchy...} & {\bf Implication}\\
  \hline \hline
   violate Kolmogorov condition &  The system cannot be simulated  by 
{\bf any} stochastic process.
 \\
 \hline
 satisfy Kolmogorov 
condition,  but
violate measurement 
independence.  
 & The system cannot be simulated by  
a  ``{\bf local}”  stochastic process, 
 \\
 \hline 
satisfy Kolmogorov 
condition and 
 measurement 
independence.  
 & The system can be simulated by a``local” 
 stochastic process. \\
   \hline
 \end{tabularx}
  \caption{Possible  behaviors of a hierarchy of probability distributions in QFT.}
\end{table}

Consider now the specific probabilistic hierarchy defined by QTP, in which $P_2(z_1, z_2) = C_2 G_{AB} R^A(z_1) R^B(z_2)$. A sufficient condition for Eq. (\ref{CauSch}) is that $G_{AB}$ is non-negative  on the Hilbert space ${\cal V}$. The existence of a negative eigenvalue of $G_{AB}$ is a necessary condition for the violation of Eq. (\ref{CauSch}). Hence, the smallest negative eigenvalue of $G_{AB}$ provides a measure of the violation of Eq. (\ref{CauSch}) that depends only on the (four-point) correlation function. 

\medskip

\section{Violation of measurement independence for scalar fields}
Next, we give examples of irreducibly quantum behavior in measurements of quantum scalar fields. In this section, we will focus on the violation of measurement independence, and we will take up non-additivity in the next section.

 Consider  a free scalar field  of mass $m$,
\bey
\hat{\phi}(x) = \int \frac{d^3k}{(2\pi)^{3/2} \sqrt{2\omega_{\bf k}}} [\hat{a}_{\bf k} e^{-ik\cdot x} + \hat{a}^{\dagger}_{\bf k} e^{ik\cdot x}],
\eey
expressed in terms of annihilation operators $\hat{a}_{\bf k}$ and creation operators $\hat{a}^{\dagger}_{\bf k}$; we  have set  $k = (\omega_{\bf k}, {\bf k})$ with $\omega_{\bf k} = \sqrt{{\bf k}^2 + m^2}$.

We assume  scalar coupling operators $\hat{C}(x)$ to the apparatus. We will also ignore any observable other than the spacetime point of detection. For an initial field state $|\psi\rangle$, the probability density $P_1(x)$ for one measurement event is
\bey
P_1(x) = C_1 \int dy \langle \psi|\hat{C}(x - \frac{1}{2}y) \hat{C}(x+ \frac{1}{2}y)|\psi \rangle R(y).
\eey
The probability density $P_2(x_1, x_2)$ for two measurement events is
\bey
P_2(x_1, x_2) = C_2 \int dy_1 dy_2 R_1(y) R_2(y) \langle \psi|{\cal T}^*[\hat{C}_1(x_1 - y_1/2)\hat{C}_2(x_2 - y_2/2)] \nonumber \\
 {\cal T} [\hat{C}_2(x_2 - y_2/2)  \hat{C}_1(x_1+y_1/2)]|\psi \rangle , \label{p2ampl}
\eey
where the two apparatuses 1 and 2 may in general be distinct: the composite operators $\hat{C}_{1,2}$ and the detection kernels $R_{1,2}$ may be different. 

Note that for scalar couplings, we can always absorb the phase $e^{i\xi\cdot x_0}$ of Eq. (\ref{tildr}) into a redefinition of the observables $x \rightarrow x - x_0$, so that $\tilde{R}$ is a positive function. Some models for the detection kernel are given in the Appendix B.

We assume an initial state that is an eigenvalue of the particle number operator. Since the composite operators are expressed in terms of creation and annihilation operators, the field state appears in the probabilities in the form of the $n$-particle reduced density matrices,
\bey
\rho_n({\bf k}_1, {\bf k}_2, \ldots, {\bf k}_n; {\bf k}'_1, {\bf k}'_2, \ldots, {\bf k}'_n) = \langle \psi|\hat{a}^{\dagger}_{{\bf k}'_1} \hat{a}^{\dagger}_{{\bf k}'_2} \ldots \hat{a}^{\dagger}_{{\bf k}'_n} \hat{a}_{{\bf k}_n}\ldots \hat{a}_{{\bf k}_2} \hat{a}_{{\bf k}_1}|\psi\rangle.
\eey

\subsection{Probability assignment}
The simplest case corresponds to   identical detectors with  $\hat{C}_1(x) = \hat{C}_2(x) = \hat{\phi}(x)$ and $R_1(x) = R_2(x) = R(x)$. We take $|\psi\rangle$ to be a state with a fixed number of particles.
Then,
\bey
P_1(x) = P_0 + C_1 \int \frac{d^3k}{(2\pi)^{3/2} \sqrt{2\omega_{\bf k}}} \frac{d^3k}{(2\pi)^{3/2} \sqrt{2\omega_{\bf k}'}} \hat{\rho}_1({\bf k}, {\bf k}') \tilde{R}\left(\frac{1}{2}(k+k')\right) e^{-i(k-k')\cdot x}, \label{dda}
\eey
where
$P_0 = C_1 \int \frac{d^3k}{ 2\omega_{\bf k} } \tilde{R}(k)$ is  the constant rate of false alarms in this set-up.
Note that in the derivation of Eq. (\ref{dda}), we used  the fact that $\tilde{R}(k) = 0$ if $k^0 < 0$.

The term $P_0$ is   background noise   to the detector, so we can drop it from the analysis.   
We define the positive operators $\hat{\Pi}_x$ on the Hilbert space of a single particle,
\bey
\langle {\bf k}|\hat{\Pi}_x|{\bf k}'\rangle = \frac{1}{2(2\pi)^3 \sqrt{\omega_{\bf k} \omega_{\bf k'}}}  \tilde{R}\left(\frac{1}{2}(k+k')\right) e^{i(k-k')\cdot x},
\eey
Then, 
we obtain $P_1(x) = C_1 Tr(\hat{\rho}_1\hat{\Pi}_x)$.

Similarly, we compute as $P_2(x_1, x_2)$ as a sum of two terms
\bey
P_2(x_1, x_2) = P^a_2(x_1, x_2) + P^b_2(x_1, x_2),
\eey
where
\bey
P^a(x_1, x_2) = \frac{C_2}{(2\pi)^6}  \int \frac{d^3k_1}{ \sqrt{2\omega_{\bf k_1}}} \int \frac{d^3k_1'}{\sqrt{2\omega_{\bf k_1'}}} \int \frac{d^3k_2}{\sqrt{2\omega_{\bf k_2}}} \int \frac{d^3k_2'}{\sqrt{2\omega_{\bf k_2'}}} \rho_2({\bf k}_1, {\bf k}_2;{\bf k}_1', {\bf k}_2')  \nonumber \\ \tilde{R}\left(\frac{1}{2}(k_1+k_1')\right)\tilde{R}\left(\frac{1}{2}(k_2+k_2')\right)e^{-i(k_1-k_1')\cdot x_1  - i (k_2 - k_2')\cdot x_2}, \label{pa3}
\\
P^b(x_1, x_2) = \frac{2C_2}{(2\pi)^6}  \mbox{Re} \int \frac{d^3k_1}{(2\pi)^3 \sqrt{2\omega_{\bf k_1}}} \int \frac{d^3k_1'}{\sqrt{2\omega_{\bf k_1'}}} \int \frac{d^3k_2}{\sqrt{2\omega_{\bf k_2}}} \int \frac{d^3k_2'}{\sqrt{2\omega_{\bf k_2'}}} \rho_2({\bf k}_1, {\bf k}_2;{\bf k}_1', {\bf k}_2')  \nonumber \\ \tilde{R}\left(\frac{1}{2}(k_1 - k_1')\right)\tilde{R}\left(\frac{1}{2}(k_2'-k_2)\right)e^{-i(k_1+k_1')\cdot x_1  + i (k_2 + k_2')\cdot x_2}
\eey

The key point here is that for $m \neq 0$ the term $P^b$ is strongly suppressed, as it involves rapid oscillations at a time-scale of order $m^{-1}$. It is also suppressed form $m = 0$, and for states with negligible support on the deep infrared. 

Dropping the term $P^b$,  we obtain
\bey
P_2(x_1, x_2) = C_2 Tr\left(\hat{\rho}_2 \hat{\Pi}_{x_1} \otimes \hat{\Pi}_{x_2}\right), \label{p2x1x2b}
\eey
Hence, in this setup, the Kolmogorov additivity condition (\ref{Kolmadd}) is satisfied---modulo the suppressed term $P_2^b(x_1, x_2)$. Physically, this makes sense, because, for this choice of coupling, particles are detected by absorption: no particle that has been measured in the first detector can be recorded by the second detector. 

Eq. (\ref{p2x1x2b}) also suggests that the violation of measurement independence in this set-up is related to the entanglement in the two-particle density matrix $\hat{\rho}_2$. In general, entanglement is not necessary. 
A special case of Eq. (\ref{p221}) is the condition in quantum optics that the second-order coherence $g^{(2)}(\tau)$ of the electromagnetic field satisfies $g^{(2)}(0) \geq 0 $. This condition is violated, for example, in experiments with anti-bunching photons \cite{QuOp}.  

\subsection{Time-of-arrival probabilities}
\begin{figure}
    \centering

 \includegraphics[width=0.7 \textwidth]{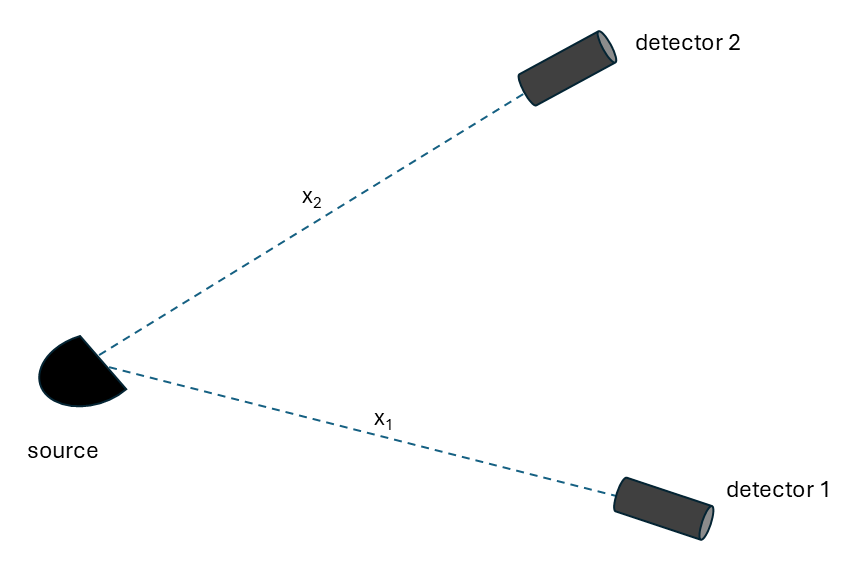}

    \caption{A two-measurement set-up: two detectors at distances $x_1$ and $x_2$ from the source. Particles effectively propagate along the line from the source to the detector.  }
    \label{fiff}
\end{figure}

To identify quantum correlations, we specialize to  a set-up that describes time-of-arrival measurements. We assume that the source is localized around ${\bf x} = 0$. In a set up with one measurement event, there is a single detector at ${\bf x}$. In a set-up with two measurements, one detector is located at ${\bf x}_1$ and the other at ${\bf x}_2$---see Fig. \ref{fiff}. We assume that the source and the detectors are motionless. This is a relativistic generalization of the setup analyzed in Ref. \cite{QTP2}.

If the size of the detectors is much smaller than the source-detector distances $|{\bf x}_1|$ and ${\bf x}_2$, then particles effectively propagate in one dimension, along the line from the source to detector. Hence, we will express each spacetime coordinate as a pair $(t_i, x_i)$, where $t_i$ is the time in the rest-frame where the detector is static and $x_i$ is the distance from the source, for $i = 1, 2$. It is convenient to take the positions of the detection events as fixed, and treat only detection times as random variables.

Hence, the probability densities $P_1$ and $P_2$ become
\bey
P_1(t) &=& \frac{C_1}{2\pi } \int \frac{dk}{ \sqrt{2\omega_k}} \frac{dk'}{\sqrt{2\omega_k'}}  \rho_1(k, k') \tilde{R}\left(\frac{1}{2}(k+k'), \frac{1}{2}(\omega_k+\omega_{k'}) \right)\nonumber \\
&\times& e^{i(k-k')x - i (\omega_k - \omega_{k'})t}, \label{p1t1}\\
P_2(t_1, t_2)  &=& \frac{C_2}{(2\pi)^2} \int \frac{dk_1}{ \sqrt{2\omega_{k_1}}} \frac{dk_2}{\sqrt{2\omega_{k_2}}}  \frac{dk_1'}{\sqrt{2\omega_{k_1'}}} \frac{dk_2'}{\sqrt{2\omega_{k_2'}}} \rho_2(k_1, k_1';k_2,k_2') \nonumber \\ &\times& \tilde{R}\left(\frac{1}{2}(k_1+k_1'), \frac{1}{2}(\omega_{k_1}+\omega_{k_1'})\right) \tilde{R}\left(\frac{1}{2}(k_2+k_2'), \frac{1}{2}(\omega_{k_2}+\omega_{k_2'})\right)  \nonumber \\
&\times& e^{i(k_1-k_1')x_1 + i(k_2-k_2')x_2 - i (\omega_{k_1}- \omega_{k_1'})t_1 - i (\omega_{k_2}- \omega_{k_2'})t_2}.
\eey
Note that the quantities $k_i$ and $k_i'$ represent here spatial momenta, and not four-momenta as in Sec. 4.1. 

 The detection probability for negative momentum particles is negligible for sufficiently large source-detector distances. It is therefore convenient to assume that the density matrices have support only on positive momenta, i.e.,  on momentum vectors in the direction from the source to the detector.  In this case there a simple normalization of those probabilities when taking time to range in the full real axis,
 \bey
 \int_{-\infty}^{\infty} dt P_1(t) = C_1 \int \frac{dk}{2k} \tilde{R}(k, \omega_k) \rho_1(k,k).
 \eey
 The quantity $\alpha(k):= \tilde{R}(k, \omega_k)/(2k)$ is the {\em absorption coefficient} of the detector, which gives fraction of incoming particles of momentum $k$ that are absorbed. We straightforwardly calculate
 \bey
 \int_{-\infty}^{\infty} dt_1  \int_{-\infty}^{\infty} dt_2 P_2(t_1, t_2) =   \frac{C_2}{2} \int dk_1 \alpha(k_1) \alpha(k_2) \rho_2(k_1,k_2; k_1',k_2').
 \eey
 We can choose $C_1$ and $C_2$ so that  $\int_{-\infty}^{\infty} dt P_1(t) =  \int_{-\infty}^{\infty} dt_1  \int_{-\infty}^{\infty} dt_2 P_2(t_1, t_2) = 1$, so that both probability densities are conditioned upon detection.
 We also re-define density matrices, post-selected along detection,
 \bey
 \tilde{\rho}_1(k,k') &=& C_1 \rho_1(k,k') \sqrt{\alpha(k)\alpha(k')} \label{ro1redef}\\
\tilde{\rho}_2(k_1,k_2; k_1',k_2') &=& C_2 \sqrt{\alpha(k_1) \alpha(k_2) \alpha(k_1') \alpha(k_2')}\rho_2(k_1, k_1';k_2,k_2') .
 \eey
 Then,
\bey
P_1(t)  = \int \frac{dk dk'}{2\pi} \tilde{\rho}_1(k,k') \sqrt{v_k v_{k'}}  L(k,k')  e^{i(k-k')x - i (\omega_k - \omega_{k'})t}, \label{p1tt}\\
P_2(t_1, t_2)  = \int \frac{dk_1dk_2dk_1'dk_2'}{4\pi^2} \tilde{\rho}_2(k_1,k_2; k_1',k_2')\sqrt{v_{k_1} v_{k'_1} v_{k_2}v_{k_2'}} \nonumber \\ \times L(k_1, k_1') L(k_2,k_2')e^{i(k_1-k_1')x_1 + i(k_2-k_2')x_2 - i (\omega_{k_1}- \omega_{k_1'})t_1 - i (\omega_{k_2}- \omega_{k_2'})t_2},
\eey
 where $v_k = k / \omega_k$ is the relativistic velocity,
  and
 \bey
L(k, k') = \frac{\tilde{R}\left(\frac{1}{2}(k+k'), \frac{1}{2}(\omega_k+\omega_{k'}) \right)}{\sqrt{\tilde{R}(k, \omega_k) \tilde{R}(k', \omega_{k'})}}
 \eey
 are the matrix elements of the {\em localization operator} $\hat{L}$. This operator is defined on the Hilbert space ${\cal H}_1$ of a single particle, and it determines the position spread of the detection record in the apparatus \cite{QTP3}. By construction, $L(k, k) = 1$. Positivity of probabilities implies that $L(k, k') \leq 1$.
 It can be shown that maximum localization is achieved for  $L(k, k') = 1$, in which case $\hat{L} = \delta (\hat{x}_{NW})$, where  $\hat{x}_{NW}$ is the Newton-Wigner position operator \cite{NWi}.

 Maximum localization corresponds to 
  an exponential form for the detection kernel ${\tilde R}(k, \omega)$,
\bey
{\tilde R}(k, \omega) = \left\{ \begin{array}{cc} A \exp \left(-\gamma_1 |k| - \gamma_0 \omega   \right), & |k|\leq \omega,
\\
0& |k| > \omega, \; \mbox{or} \; \omega <0, \end{array}\right.
\eey
for some constants $\gamma_1, \gamma_0 \geq 0$, and $A > 0$.

We can also write
\bey
P_1(t) = Tr \left( \hat{U}(x, t) \hat{\tilde{\rho}}_1 \hat{U}^{\dagger}(x, t) \sqrt{\hat{v}} \hat{L} \sqrt{\hat{v}}\right),
\eey
where $\hat{U}(x, t) = e^{i\hat{p}x - i \hat{h}t}$ is a unitary operator implementing spacetime translation, $\hat{p}$ is the momentum operator, $\hat{h} = \sqrt{m^2+\hat{p}^2}$ is the Hamiltonian, and $\hat{v} = \hat{p}\hat{h}^{-1}$ is the relativistic velocity operator. This means that 
\bey
P_1(t) &=& Tr[\hat{\tilde{\rho}}_1 \hat{\Pi}_x(t)] \\
P_2(t_1, t_2) &=& Tr[\hat{\tilde{\rho}}_2 [\hat{\Pi}_{x_1}(t_1)\otimes \hat{\Pi}_{x_2}(t_2)]],
\eey
where the POVM $\hat{\Pi}_x(t)$ are defined by
\bey
\hat{\Pi}_x(t) = \sqrt{\hat{v}} \hat{U}^{\dagger}(x, t) \hat{L}\hat{U}(x, t)  \sqrt{\hat{v}}. \label{pxtt}
\eey
For maximum localization, the POVM (\ref{pxtt}) was first derived in \cite{Leon} as a relativistic generalization of Kijowski's POVM \cite{Kij} for the time of arrival.

%\bey
%\langle k' |\hat{\Pi}_x(t)|k\rangle  = \sqrt{v_k v_{k'}}  e^{i(k-k')x - i (\omega_k - \omega_{k'})t}.
%\eey

\subsection{Violation of measurement independence}

An entangled state $\hat{\tilde{\rho}}_2$ may lead to probabilities that violate measurement independence, as expressed by Eqs. (\ref{pepe}) and (\ref{quant2}).
To see this, we  
  consider a pure state
\bey
\psi(k_1, k_2) = \frac{1}{\sqrt{2 (1 + |\epsilon|^2)}} \left[\phi_1(k_1) \phi_2(k_2) +\phi_1(k_2) \phi_2(k_1)\right],
\eey
for generic single particle-states $\phi_1$ and $\phi_2$; $\epsilon = \int dk \phi_1(k) \phi_2^*(k)$.
Assuming  maximum localization, we find  
\bey
P_1(t) &=& \frac{1}{2(1+|\epsilon|^2)} \left[ |{\cal A}_1(t)|^2   + |{\cal A}_2(t)|^2 + \epsilon {\cal A}_1(t) {\cal A}_2^*(t) + \epsilon^* {\cal A}^*_1(t) {\cal A}_2(t)  \right],\hspace{0.6cm} \\
P_2(t_1, t_2) &=& \frac{1}{2(1+|\epsilon|^2)} \left| {\cal A}_1(t_1)   {\cal A}_2(t_2) +    {\cal A}_1(t_2)   {\cal A}_2(t_1)   \right|^2,
\eey
where
\bey
{\cal A}_i(t) = \int_{-\infty}^{\infty} dk \phi_i(k)\sqrt{v_k} e^{ikx - i \omega_kt}. \label{Amplc}
\eey
Next, we evaluate 
\bey
w(t) = P_1(t)^2 - P_2(t, t) = \frac{1}{4} \left[|{\cal A}_1(t)|^4 + |{\cal A}_2(t)|^4 - 6 |{\cal A}_1(t)|^2 |{\cal A}_2(t)|^2 \right]. \label{wtf}
\eey
for $\epsilon = 0$. Eq. (\ref{pepe}) is violated for 
\bey
|{\cal A}_1(t)/{\cal A}_2(t)|\notin [\sqrt{3-2\sqrt{2}},\sqrt{3+2\sqrt{2}}] \simeq [0.41, 2.41]. \label{bounds1}
\eey
 
We also define  
\bey
G(t_1, t_2) :&=&  P_2(t_1, t_2) -  \sqrt{P_2(t_1, t_1) P_2(t_2, t_2)} \nonumber \\
&=& \frac{1}{2}\left(|{\cal A}_1(t_1)  {\cal A}_2(t_2)|  -  |{\cal A}_1(t_2)  {\cal A}_2(t_1)|   \right)^2  \nonumber \\
&-& 2 |{\cal A}_1(t_1)  {\cal A}_2(t_2) {\cal A}_1(t_2)  {\cal A}_2(t_1)| \sin^2\Theta(t_1, t_2),
\eey
where  $\Theta := \frac{1}{2}(\theta_{11} + \theta_{22} - \theta_{12} - \theta_{21})$, with $\theta_{ij} :=  \mbox{arg} {\cal A}_i(t_j)$. Eq. (\ref{quant2}) is violated when  $G(t_1, t_2) > 0$, or, equivalently, when
\bey
\sin^2\Theta < \frac{1}{4}\left[ \frac{|{\cal A}_1(t_1)   {\cal A}_2(t_2)|}{|{\cal A}_1(t_2)   {\cal A}_2(t_1)|} +  \frac{|{\cal A}_1(t_2)   {\cal A}_2(t_1)|}{|{\cal A}_1(t_1)   {\cal A}_2(t_2)|} - 2     \right]. \label{ineqa}
\eey
Inequality (\ref{ineqa}) is always satisfied for $\Theta$ in a neighbourhood of $\Theta = 0$. Furthermore, it is violated for all $\Theta$, if the ratio
\bey
\eta = \frac{|{\cal A}_1(t_1)   {\cal A}_2(t_2)|}{|{\cal A}_1(t_2)   {\cal A}_2(t_1)|}  \notin [3 - 2\sqrt{2}, 3 + 2\sqrt{2}]. \label{bounds2}
\eey
Eq. (\ref{bounds2}) is compatible with Eq. (\ref{bounds1}).

The amplitudes ${\cal A}_i(t)$ typically oscillate rapidly with the energy scale of the initial state. For example, if $\phi_i(k)$ is strongly peaked around $k = p_i$, we evaluate the integral  (\ref{Amplc}) by expanding the phase around $k = p$. Then,  we obtain
\bey
{\cal A}_i(t) = v_{p_i} e^{i\omega_{p_i}t} \tilde{\phi}_i(x - v_{p_i}t), \label{approxam}
\eey
where $\tilde{\phi}_i$ is the inverse Fourier transform of $\phi_i$.

For a quantitative estimate, take $p_1 = p_2 = p$, and use for $\phi_1$ and $\phi_2$ identical Gaussians with their centers separated  by distance $a$ for $\phi_1$ and $\phi_2$ : 
 \bey
 \tilde{\phi}_1(x) = (2\pi\sigma^2)^{-1/4} \exp[-x^2/(4\sigma^2)+ipx], \;\;\; \tilde{\phi}_2(x) = \phi(x - a). \nonumber
 \eey 
  We straightforwardly evaluate $\eta = \exp\left[\dfrac{av_p \Delta t}{2\sigma^2}\right]$, where $\Delta t = t_2 - t_1$. Then,  the violation of Eq. (\ref{quant2}) is guaranteed to occur for $|\Delta t| \gtrapprox 2.5 \dfrac{\sigma^2}{av_p}$.
Furthermore, for $a \gtrapprox 5 \sigma$, $\langle \phi_1|\phi_2\rangle << 1$, and Eq. (\ref{wtf}) applies. It is straightforward to show that Eq. (\ref{pepe}) is violated for $\left|t -  \dfrac{L-\dfrac{1}{2}a}{v_p}\right| \gtrapprox 1.76 \dfrac{\sigma^2}{av_p}$.

\begin{figure}
 \includegraphics[width=8.6cm]{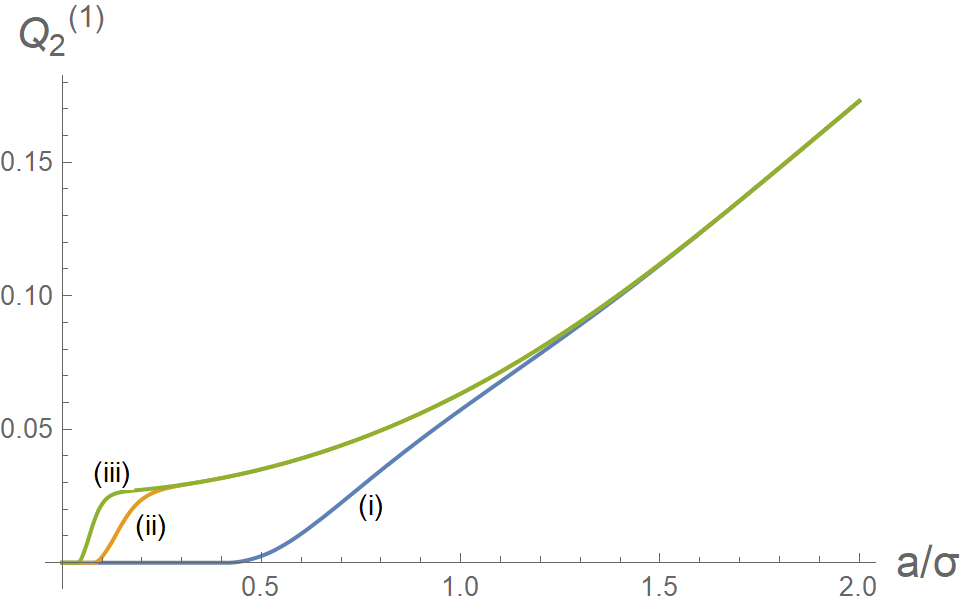}\label{q11fig}
     \caption{The measure $Q^{(1)}_2$ of Eq. (\ref{qn1}) for $m = 0$ as a function of $a/\sigma$ and for different values of the source-detector distance $x$: (i) $a/x = 0.1$, (ii) $a/x = 0.02$, and (iii) $a/x = 0.01$. }
\end{figure}

In Fig. (\ref{q11fig}), we plot the non-classicality measure $Q^{(1)}_2$ of Eq. (\ref{qn1}) as a function of $a/\sigma$, for $m = 0$. As expected, this increases with $a$, that is, with the size of the superposition. 

Within the approximation (\ref{approxam}), we can evaluate the non-classicality measure $Q^{(2)}_2 = \int dt_1 dt_2 G(t_1, t_2)$ analytically. We find that
\bey
Q^{(2)}_2 = 1 - e^{-\frac{a^2}{8\sigma^2}},
\eey
that is, $Q^{(2)}_2$ increases with the ratio $a/\sigma$, and converges rapidly to the maximum value $1$.

\section{Violation of Kolmogorov additivity}

In this section, we consider a model for scalar field measurements that allows us to probe the violation of Kolmogorov additivity.

\subsection{The measurement model}

We will again consider measurements of a single scalar field $\hat{\phi}(x)$, but now we will consider different couplings to the apparatus. For a single measurement event, we will employ the composite operator  $\hat{C}_1(x) = : \hat{\phi}(x)^2:$ with detection kernel $R_1(x)$. For two measurement events, the first apparatus will again be described by $\hat{C}_1$ and $R_1$, but for the second apparatus we will employ   $\hat{C}_2(x) = \hat{\phi}(x)$ with detection kernel $R_2$. The key point here is that the first apparatus records particles either by scattering or by double absorption. A particle that has been scattered can be absorbed by the second detector, hence, in this case a violation of Kolmogorov additivity is possible.

The calculation of the detection probabilities  is straightforward if long. For a single detection event, two terms survive in the probability density $P_1$; for two detection events four terms survive in the probability density $P_2$. Their explicit expressions are given in the Appendix. The forms of these terms are expressed

\begin{figure}
    \centering

 \includegraphics[width=1 \textwidth]{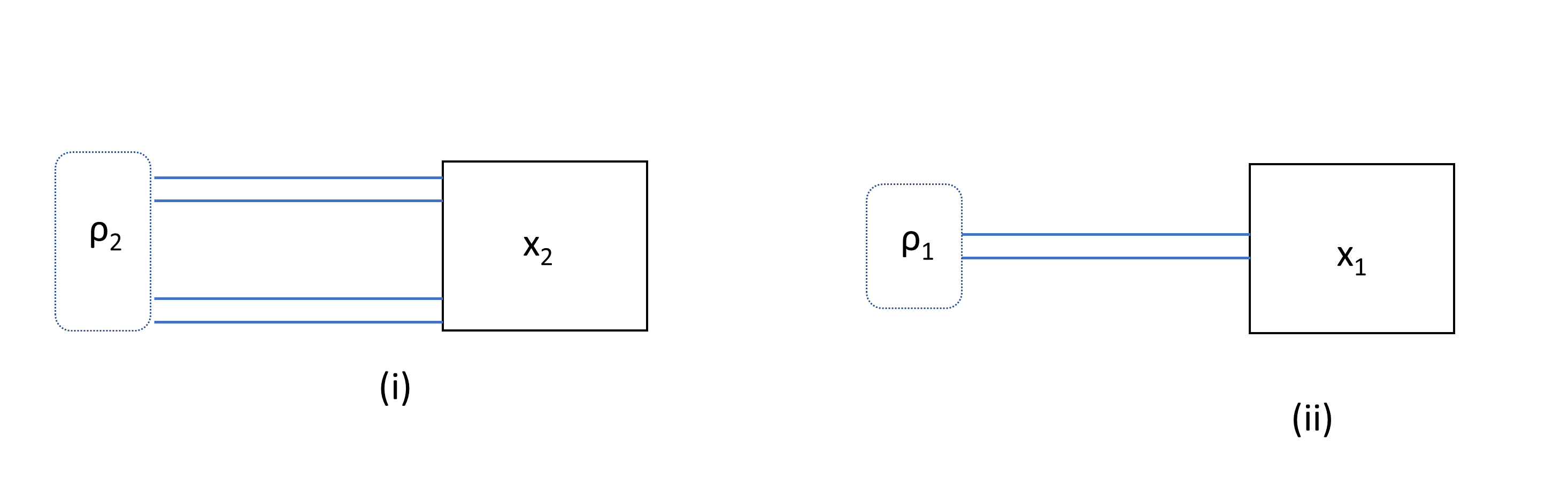}

    \caption{ Diagrammatic form of the two terms that appear in the probability density $P_1(x)$ for a single detection event. }
    \label{diagrams0}
\end{figure}

\begin{figure}
    \centering

 \includegraphics[width=1 \textwidth]{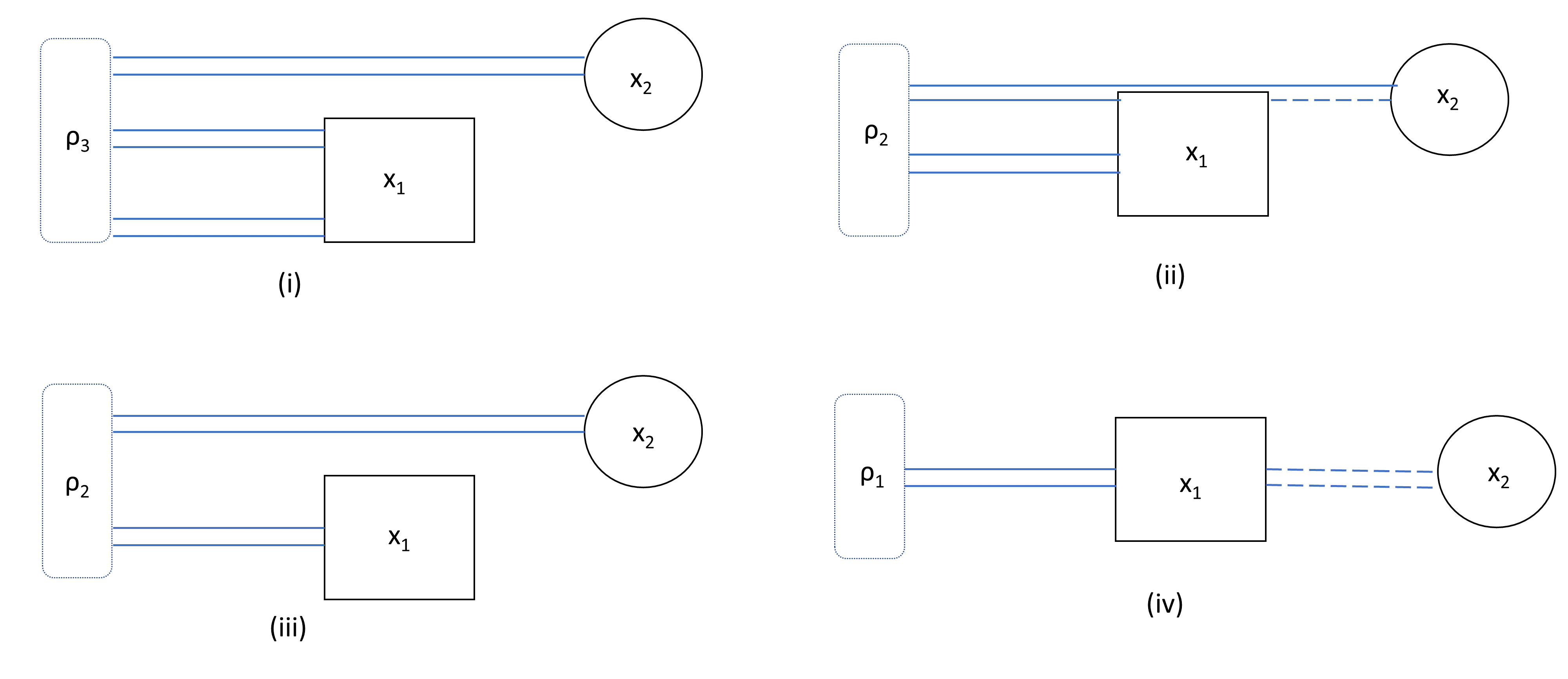}

    \caption{  Diagrammatic form of the four terms that appear in the probability density $P_2(x_1, x_2)$ for two  detection events. }
    \label{diagrams}
\end{figure}

Four terms survive, and their explicit expressions are given in the Appendix. Their form is represented diagrammatically in Fig. \ref{diagrams0} for $P_1$ and in
 Fig. \ref{diagrams} for $P_2$. Each particle corresponds to two lines, one for each argument of its density matrix. Solid lines correspond to incoming particles, dashed lines correspond to particles that propagate after interaction with the first apparatus.

The term (i) in $P_1(x)$ corresponds to simultaneous absorption of two particles, and as such  it involves information from the two-particle reduced density matrix $\rho_2$. The term (ii) corresponds to the detection of one particle through scattering, and it involves information in the one-particle reduced density matrix.

\begin{itemize}
\item Terms of type (i) involve contributions to the joint detection probability from triplets of particles, two particles being detected by the first apparatus, and one particle by the second. It involves information contained in the three-particle reduced density matrix, $\rho_3$.
 
\item Terms of type (ii) involve contributions from pairs of particles. One particle is detected by the first detected, and another particle is determined by the second detector, but while also receiving a contribution from particle scattering in the first detector. The information is contained in the two-particle reduced density matrix $\rho_2$.

\item Terms of type (iii) are also determined by $\rho_2$. They correspond to two distinct events at the two detectors. Their contribution to the probability density is of a  form similar to Eq. (\ref{pa3}).

\item Finally, terms of type (iv) are determined by the one-particle reduced density matrix $\rho_1(k, k')$. They correspond to a particle being detected through scattering in the first apparatus, and then detected in the second apparatus.
\end{itemize}

Hence, the probability distribution for two detection events is a sum of four terms $P_2(x_1, x_2) = \sum_{a=1}^4 P_2^{(a)}(x_1, x_2)$, where $P_2^{(a)}(x_1, x_2)$ corresponds to the $a$-th diagram of Fig. \ref{diagrams}. For a single-particle state, all $n$-particle reduced density matrices vanish except for $\rho_1$. Hence, only diagram (ii) of Fig. \ref{diagrams} contributes to $P_1$, and only diagram (iii) of Fig. \ref{diagrams} contributes to $P_2$.

\subsection{Probability assignment}

\begin{figure}
    \centering

 \includegraphics[width=0.7 \textwidth]{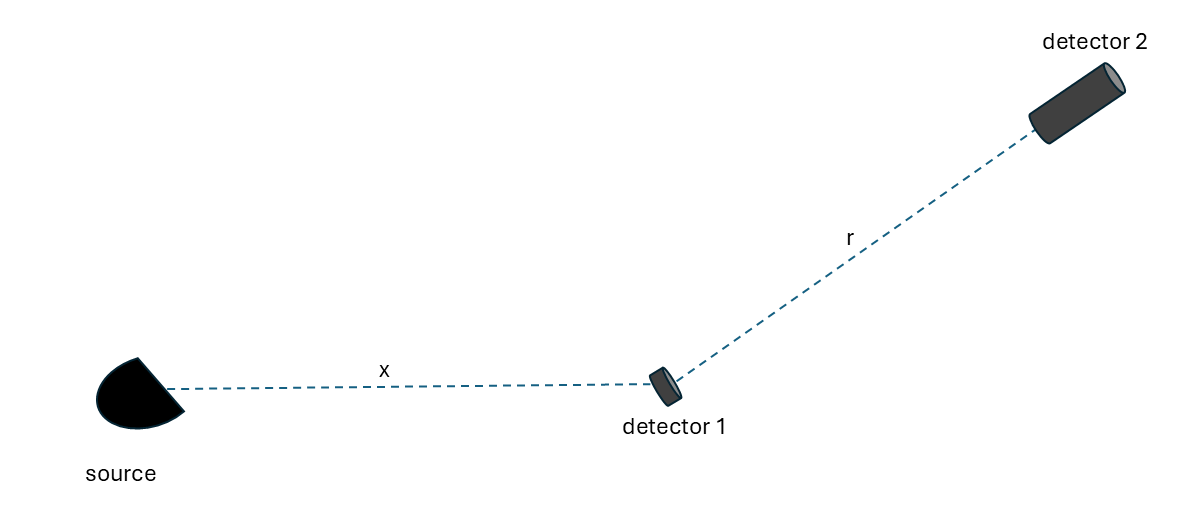}

    \caption{A set-up in which a particle is recorded by the first detector through scattering, and it is subsequently recorded by the second detector. }
    \label{renfeea}
\end{figure}

We consider a set-up with two detectors, in which a particle scatters off the first detector, after leaving a measurement record, and then propagates to a second detector\footnote{The presence of a measurement record for the time of interaction in the locus of the first scattering is the key difference of this setup from the ones encountered in textbook treatments of scattering.}. The direction of scattering is a random variable; however, given a fixed position of the source and the two detectors, only particles scattered within a small solid angle along the line connecting the two detectors leave a record in both detectors. Hence, when working on the sub-ensemble of particles that leave two detection records, the scattering angle is fixed by the geometry of the configuration\footnote{We can include the scattering angle as a random variable, by considering an extended spherical array of detectors covering the locus of the first detector. However, this is not necessary for the purposes of demonstrating violation of Kolmogorov additivity.}. Then, the only random variables are the times of detection. We denote by $t$ the detection time at detector $1$ and by $t + \tau$ the  time of the second direction. We also denote the distance between the source and the first detector by $x$ and the distance between the two detectors by $r$---see Fig. \ref{renfeea}.

 The probability density $P_1(t)$ is given by an identical expression with Eq. (\ref{p1t1}), 
\bey
P_1(t) = \frac{C_1}{2\pi } \int \frac{dk}{ \sqrt{2\omega_k}} \frac{dk'}{\sqrt{2\omega_k'}}  \rho_1(k, k') \tilde{R}_1'\left(\frac{1}{2}(k+k'), \frac{1}{2}(\omega_k+\omega_{k'}) \right) e^{i(k-k')x - i (\omega_k - \omega_{k'})t}, \label{P1t1b}
\eey
the only difference being that the detection kernel is given by
\bey
\tilde{R}_1'(k, \omega) = \int \frac{dp}{\omega_p} \tilde{R}_1(k - p, \omega - \omega_p). \label{r1prime}
\eey
As in Sec. 4.2, we absorb $C_1$ in a redefinition $\hat{\rho}_1 \rightarrow \hat{\tilde{\rho}}_1$ via Eq. (\ref{ro1redef}). Then, we obtain Eq. (\ref{p1tt}) for the probability density $P_1(t)$ conditioned upon detection, with localization operator
  \bey
 L_1(k, k') = \frac{\tilde{R}'\left(\frac{1}{2}(k+k'), \frac{1}{2}(\omega_k+\omega_{k'}) \right)}{\sqrt{\tilde{R}'(k, \omega_k) \tilde{R}'(k', \omega_{k'})}} \label{loc1}
 \eey
 The probability density for two detections at time $t$ and $t + \tau$, for $\tau > 0$, is 
\bey
P_2(t, \tau) &=& C_2 \int \frac{dkdk'}{4\pi\sqrt{\omega_k \omega_{k'}}}  \rho_1(k, k')  e^{i (k-k') x - i (\omega_k - \omega_{k'})t} \hspace{2.5cm}
\nonumber \\
&\times& \int \frac{dqdq'}{4\pi \omega_{q}\omega_{q'}}    \tilde{R}_1\left(\frac{1}{2}(k+k'-q - q'), \frac{1}{2} (\omega_k + \omega_{k'} - \omega_q  - \omega_{q'})\right) \nonumber \\
&\times&  \tilde{R}_2\left(\frac{1}{2}(q+q'), \frac{1}{2}(\omega_q + \omega_{q'})\right) 
e^{ i (q-q')r - i (\omega_q - \omega_{q'})\tau}, \label{p212}
\eey
where $C_2$ is a normalization constant.

The total probability of detection is
\bey
\int dt d\tau P_2(t, \tau) = C_2 \int \frac{dk}{2k} \alpha_2*\tilde{R}'_1(k, \omega_k),
\eey
where $\alpha_2*\tilde{R}'_1$ is an involution of $\alpha_2(q) = \tilde{R}_2(q, \omega_q)/(2q) $ with $\tilde{R}_1'(k, \omega)$,
\bey
\alpha_2*\tilde{R}'_1(k, \omega) = \int \frac{dq}{\omega_q} \alpha_2(q) \tilde{R}_1(k - q, \omega - \omega_q).
\eey 
Hence, the total absorption coefficient for this experiment is $\alpha_{1,2}(k) =  \alpha_2*\tilde{R}'_1(k, \omega_k)/(2k)$. We can define an initial state conditioned on detection
\bey
\tilde{\rho}_1(k, k') = C \rho_1(k, k') \sqrt{\alpha_{1,2}(k) \alpha_{1,2}(k')} ,  \label{staten4}
\eey
so that $\int dt d \tau  P_2(t, \tau) = 1$. 

We therefore obtain
\bey
P_2(t, \tau) = \int \frac{dkdk'}{2\pi} \sqrt{v_kv_{k'}} \tilde{\rho}_1(k,k')e^{i (k-k') x - i (\omega_k - \omega_{k'})t} \nonumber \\
\times \int \frac{dq dq'}{2\pi} \sqrt{v_qv_{q'}} S(k,q; k', q') L_2(q, q') e^{ i (q-q')r - i (\omega_q - \omega_{q'})\tau}, \label{p2t1t2}
\eey
where $\hat{L}_2$ is the localization operator for the second detector, and $S(k,q; k', q')$ are the matrix elements of an operator $\hat{S}$ on ${\cal H}_1 \otimes {\cal H}_2$
\bey
\langle k, q| \hat{S}|k', q'\rangle = \frac{\tilde{R}_1\left(\frac{1}{2}(k+k'-q - q'), \frac{1}{2} (\omega_k + \omega_{k'} - \omega_q  - \omega_{q'})\right)}{\sqrt{\omega_q \omega_{q'} \; \alpha_2*\tilde{R}_1'(k, \omega_k)\;  \alpha_2*\tilde{R}_1'(k', \omega_{k'})}}\sqrt{\alpha_2(q) \alpha_2(q')}, \label{reducop}
\eey 
For reasons that will be made apparent shortly, we will call $\hat{S}$ the {\em reduction operator}. This operator is self-adjoint, and invariant under the partial transposition $\langle k, q| \hat{S}|k', q'\rangle \rightarrow \langle k, q'| \hat{S}|k', q\rangle$.

The second partial trace of $\hat{S}$ is a modified version $\hat{L}_1^*$ of the localization operator $\hat{L}_1$, obtained by replacing $\tilde{R}_1'$ by $\alpha_2*\tilde{R}_1'$ in Eq. (\ref{loc1}),
\bey
Tr_2 \hat{S} = \hat{L}_1^*.
\eey
This change in the localization operator is due to the normalization of probabilities, and not an effect of the action of the second detector upon the first.
In Eq. (\ref{p2t1t2}), we  consider a statistical sub-ensemble of particles recorded at both detectors, so quantities pertaining to the first measurement have to be ``averaged" with respect to the detection statistics of the second measurement. If the absorption coefficient of the second detector is constant, then $\hat{L}_1 = \hat{L}_1^*$. 

We also note that 
\bey
\int dq \langle k, q| \hat{S}|k, q\rangle = 1, \label{normmm}
\eey
The operator $\hat{\sigma}_k$ defined by 
\bey
\langle q|\hat{\sigma}_k|q'\rangle = \langle k, q| \hat{S}|k, q'\rangle \label{sigmaop}
\eey
 is self-adjoint. By Eq. (\ref{normmm}), $\hat{\sigma}_k$ is a density matrix for an individual particle. Its diagonal elements $\langle q|\hat{\sigma}_k|q\rangle$  define the probability that a particle with incoming momentum $k$ is scattered towards the direction $q$. They are proportional to the differential scattering cross-section along the line that connects the two detectors.

\subsection{Relativistic state reduction}
The probability density $P_2(t, \tau)$ of Eq. (\ref{p2t1t2}) satisfies the Kolmogorov condition when integrating over the time $\tau$ of the second measurement,
\bey
\int d\tau P_2(t, \tau) = P_1(t),
\eey
modulo the change $\hat{L}_1 \rightarrow \hat{L}_1^*$ for the localization operator, and the corresponding change (\ref{staten4}) in $\hat{\rho}_1$, to accommodate for post-selection in the statistical ensembles. 
In contrast, the Kolmogorov condition is violated when integrating $P_2(t, \tau)$ with respect to $t$.
To see this, we write 
\bey
P(t, \tau) = \frac{1}{4\pi^2} Tr \left[\left(  \sqrt{\hat{v}} \hat{U}^{\dagger}(r, \tau) \hat{L}_2 \hat{U}(r, \tau) \sqrt{\hat{v}} \otimes \hat{I}\right)    \right. \nonumber \\
\left. \times \hat{S}\left( \sqrt{\hat{v}}\hat{U}(x, t ) \hat{\rho}_1 \hat{U}^{\dagger}(x, t)  \sqrt{\hat{v}} \otimes \hat{I}\right)    \right].
\eey
The conditional probability density $P(\tau|t)$ of a second detection at time $\tau$ after the first detection,  given that the first detection occurred at time $t$ is given by 
\bey
P(\tau|t) = \frac{P_2(t, \tau)}{P_1(t)}= \frac{1}{2\pi} Tr\left[\sqrt{\hat{v}}\hat{U}(r, \tau) \hat{\rho}_1^{(x, t)}\hat{U}^{\dagger}(r, \tau) \sqrt{\hat{v}} \hat{L}_2 \right],
\eey
where 
\bey
\hat{\rho}_1^{(x, t)} = \frac{1}{2\pi P_1(t)}  \;   Tr_1 \left[\hat{S}\left( \sqrt{\hat{v}}\hat{U}(x, t) \hat{\tilde{\rho}}_1 \hat{U}^{\dagger}(x, t)  \sqrt{\hat{v}} \otimes \hat{I}\right)    \right] \;  
\eey 
is the effective density matrix for the particle exiting the first detector after being recorded. It is straightforward to show that $Tr \hat{\rho}_1^{(x, t)} = 1$.
The transformation $\hat{\tilde{\rho}}_1 \rightarrow \hat{\rho}_1^{(x, t)}$ is a quantum-state reduction rule for a particle recorded at the spacetime point $(x, t)$, a fact that justifies the name of the operator $\hat{S}$.  

For an  initial state that is almost monochromatic at momentum $k_0$, we can approximate $\hat{S} \simeq L_1^* \otimes \hat{\sigma}_{k_0}$, so that 
\bey
\hat{\rho}_1^{(x, t)} \simeq \hat{\sigma}_{k_0},
\eey
that is, the effective density matrix carries no  memory of the initial state, except for momentum. Otherwise, it is fully determined by the properties of the apparatus.

We emphasize that in the QTP analysis, the reduction rule follows from the form of the joint probability $P_2$, which ultimately derives from the decoherent histories probability assignment. Hence, the reduction rule is a derivative and not a fundamental notion. This is why the reduction rule in the first detector depends on the normalization of probabilities in the second detector. There is no action backwards in time, simply a redefinition of the relevant statistical sub-ensemble due to postselection.
 For other derivations of the reduction rule in analogous setups, see \cite{QTP2, GGM22, FJR22}.

When integrating $P_2(t, \tau)$ with respect to $t$, we construct a probability density for $\tau$ in the second detector, with an initial state obtained from a non-selective measurement in the first detector
\bey
\hat{\rho}_{1}^{ns} = \int dt P_1(t) \hat{\rho}_1^{(x, t)}.
\eey
 We straightforwardly calculate
 \bey
 \rho_{1}^{ns}(q, q') = \int dk \tilde{\rho}_1(k, k) S(k,q; k, q'),
 \eey
 or, equivalently,
 \bey
 \hat{\rho}_{1}^{ns} = \int dk \tilde{\rho}_1(k, k) \hat{\sigma}_k.
\eey
Hence, the state of the statistical ensemble after measurement is determined by the properties of the detector, as expressed by the density matrices $\hat{\sigma}_k$. 
All information about the off-diagonal elements of $\hat{\rho}_1$ in the momentum basis is lost. The diagonal elements of $\hat{\rho}_1$ provide the weight of 
 the contribution of $\hat{\sigma}_k$ of different momenta in the final state.
 
 The measure of the violation of Kolmogorov inequality is the statistical distance $w_1$ between $\tilde{P}(\tau) = \int dt P_2(t, \tau)$ and $P_1(\tau)$. This equals
 \bey
 w_1 = \sup_A \left|\int d\tau  \chi_A(\tau) [\tilde{P}(\tau) - P_1(\tau)]\right| = \sup_A \left| Tr \left[ (\hat{\rho}_1^{ns} - \hat{\tilde{\rho}}_1)\hat{\Pi}_A\right]\right|. \label{w112}
 \eey
where the supremum is over all subsets $A$ of the real line, and we wrote $\hat{\Pi}_A = \int \hat{\Pi}_{\tau} \chi_A(\tau)$. The right-hand-side of Eq. (\ref{w112}) is smaller than $\sup_{\hat{E}} \left| Tr \left[ (\hat{\rho}_1^{ns} - \hat{\tilde{\rho}}_1)\hat{E}\right]\right|$, where $\hat{E}$ ranges over all positive operators with $\hat{E} \leq \hat{I}$. The latter quantity coincides with $\frac{1}{2} Tr|\hat{\rho}_1^{ns} - \hat{\tilde{\rho}}_1|$ \cite{Wilde}, so we obtain an upper limit 
\bey
w_1 \leq \frac{1}{2} Tr|\hat{\rho}_1^{ns} - \hat{\tilde{\rho}}_1|.
\eey

\section{Conclusions}

The main result of this paper is the analysis of quantum resources in QFT. We showed that the QTP description of measurements enables the construction of probabilistic hierarchies that probe the information content in the QFT hierarchy of unequal-time correlation functions. In this setting, we analyzed quantum resources, in particular ones that refer to the compatibility between different levels of the probabilistic hierarchy. We identified two types of irreducibly quantum behavior. The first is associated with the failure of Kolmogorov additivity, and the second with the failure of measurement independence.  Then, we considered specific examples of those resources, by focusing on measurements in which the main random variable is the detection times of relativistic particles. A by-product of this analysis was the derivation of a reduction rule for relativistic particles that are recorded---through scattering---at a spacetime point $x$. 

This paper is guided by the belief that the unequal time correlation functions are the essential physical content of QFT. This implies that any relativistic approach to quantum information must focus on the properties of the hierarchy of such correlation functions, and not on the properties of the quantum state as in non-relativistic physics. The correlation functions contain information that pertain both to the initial (and occasionally final) conditions of a QFT and to dynamics, and in a relativistic setting it is impossible to separate them\footnote{There are mathematical reasons for this. For example, to select one out of the infinite representations of the canonical (anti)commutation relations for a quantum field, one requires that an operator corresponding to the classical field Hamiltonian is defined on the relevant Hilbert space.}.\

 We believe that the information measures considered here must be included in any discussion of information balance in QFT. This is particularly relevant for the discussion of information loss in black-hole evaporation. In previous work, we have shown that multi-detector measurements (carried out at finite times) reveal that the Hawking radiation is not thermal or close to thermal \cite{AnSav19, AnSav19b}. Only the probabilities associated to single-event measurements are thermal.

In this paper, we quantified quantum behavior using simple $L^1$ and supremum norms for the violation of classicality conditions. Entropic measures might be more appropriate for applications to non-equilibrium QFT. We recall that the level $n=1$ of the probabilistic hierarchy is the level at which the Boltzmann equation---and the thermodynamic entropy---is defined \cite{CH08}. A relation between thermodynamic entropy and entropic measures of quantum correlations from the higher field correlations could be important for understanding the origins of macroscopic irreversibility. 

Our longer term aim is to define quantum resources with reference solely to the hierarchy of correlation functions, and not in relation to specific measurement set-ups, as in this paper. This would provide a unified treatment of quantum resources in QFT that would bring together Kolmogorov non-additivity, violation of measurement independence (as defined here), and Bell-type non-locality. To this end, it is necessary to have a general characterization of all measurements admissible in QFT, i.e., to identify all POVMs that are compatible with relativistic causality. At the moment, this seems a formidable task.

We expect that the irreducible quantum behavior identified in this paper is experimentally accessible. For example, we noticed that the violation of the inequality (\ref{p221}) is already established in anti-bunching experiments for photons. In a future publication, we will analyse how the detection-time correlations of Sec. 4 can be assessed experimentally, possibly in quantum optics experiments in space that involve long baselines \cite{DSQL}. The violation of Kolmogorov additivity is more intricate, but it may be possible to employ variations of setups  that test macrorealism through the violation of the Leggett-Garg inequality \cite{LeGare}.

Finally, we note that  in our analysis,  the state-reduction rule arises as a secondary consequence of the probability rule for multi-detector measurements. It is certainly not a physical process, and this is why it may carry information from subsequent measurements, when we condition our statistical ensemble upon detection. This treatment of reduction 
 is in accordance with ideas first expressed by Wigner \cite{WH65}, and made explicit in the decoherent histories program. Nonetheless, our methodology and the specific expressions for reduction derived here may be useful to the program of dynamical reduction \cite{GhBa}, since these expressions are derived from a fully-fledged relativistic quantum theory.

\section*{Acknowledgements}
Research was supported by   grant  JSF-19-07-0001 from the Julian Schwinger Foundation. We thank Bei Lok Hu for discussions and suggestions on an early version of this manuscript.

\begin{appendix}
\section{The two-event probability density of Sec. 5}
The probability for one measurement event with  $\hat{C}(x) = : \hat{\phi}(x)^2:$ consists of two terms, each corresponding to one of the two diagrams in Fig. (\ref{diagrams0}),
\bey
P_1^{(i)}(t) &=& \int \frac{d^3q d^3q' d^3k  d^3k'}{(2\pi)^{6} 4 \sqrt{\omega_{{\bf q}}\omega_{{\bf q}'} \omega_{{\bf k}}\omega_{{\bf k}'}}} \, \rho_2({\bf q}, {\bf k}; {\bf q}',  {\bf k}') \nonumber \\
&\times&\left[ \tilde{R}_1\left(\frac{k+k'+q+q'}{2} \right) e^{-i(k+q-k'-q')\cdot x} + \tilde{R}_1\left(\frac{k+k' - q - q'}{2} \right) e^{-i(k+q-k'-q')\cdot x}\right], \nonumber  \\
P_1^{(ii)}(t) &=& 4 \int \frac{d^3k d^3k'}{(2\pi)^{3} 2 \sqrt{\omega_{{\bf k}}\omega_{{\bf k}'}} }\, \rho_1({\bf k}; {\bf k}') \tilde{R}_1'\left( \frac{k+k'}{2}\right) e^{-i(k-k')\cdot x}.
\eey

The  probability density $P_2(x_1, x_2)$ for two measurement events for $\hat{C}_1(x) = : \hat{\phi}(x)^2:$ and $\hat{C}_2(x) = \hat{\phi}(x)$ receives four contributions, each corresponding to the four diagrams of Fig. \ref{diagrams}, 
\bey
P_2^{(i)} &=& \int \frac{d^3q_1 d^3q_2 d^3k d^3q_1' d^3q_2' d^3k'}{(2\pi)^{9} 8 \sqrt{\omega_{{\bf q}_1}\omega_{{\bf q}_2}\omega_{{\bf q}'_1} \omega_{{\bf q'}_2}\omega_{{\bf k}}\omega_{{\bf k}'}}} \, \rho_3({\bf q}_1, {\bf q}_2, {\bf k}; {\bf q}_1', {\bf q}_2, {\bf k}')
\nonumber \\
&\times& \tilde{R}_1\left(\frac{q_1+q_2+q_1'+q_2'}{2}\right) \tilde{R}_2\left(\frac{k+k'}{2}\right) e^{-i(q_1+q_2-q_1'-q_2')\cdot x_2 - i (k - k') \cdot x_2},
\\
P_2^{(ii)} &=&4 \int \frac{d^3q d^3q' d^3k  d^3k'}{(2\pi)^{6} 4 \sqrt{\omega_{{\bf q}}\omega_{{\bf q}'} \omega_{{\bf k}}\omega_{{\bf k}'}}} \, \rho_2({\bf q}, {\bf k}; {\bf q}',  {\bf k}') e^{-i(q-q')\cdot x_1} \nonumber \\
&\times&   \left[F(k'-q'-q,k|x_2 - x_1) e^{ik' \cdot x_1 - i k \cdot x_2}  + F(k'-q'-q,k'|x_2 - x_1) e^{-ik' \cdot x_1 + i k \cdot x_2}  \right] \nonumber
\\
P_2^{(iii)} &=&4 \int \frac{d^3q d^3q' d^3k  d^3k'}{(2\pi)^{6} 4 \sqrt{\omega_{{\bf q}}\omega_{{\bf q}'} \omega_{{\bf k}}\omega_{{\bf k}'}}} \, \rho_2({\bf q}, {\bf k}; {\bf q}',  {\bf k}') \nonumber \\
&\times& \tilde{R}_1' \left(\frac{q+q'}{2}\right) \tilde{R}_2\left(\frac{k+k'}{2}\right) e^{- i (q - q')\cdot x_1 - i (k - k') \cdot x_2},
\\
P_2^{(iv)} &=&4 \int \frac{d^3q d^3q'}{(2\pi)^{3} 2 \sqrt{\omega_{{\bf q}}\omega_{{\bf q}'}} }\, \rho_1({\bf q}; {\bf q}') G\left(\frac{1}{2} (q + q')|x_2 - x_1\right) e^{-i(q-q')\cdot x_1}
\eey
where
\bey
F(q,q'|x) &=& \int \frac{d^3p}{(2\pi)^3 2 \omega_{\bf p}} \tilde{R}_1\left(\frac{p+q}{2}\right) \tilde{R}_2\left(\frac{p+q'}{2}\right) e^{-i p \cdot(x_2 - x_1)}, \\
\tilde{R}_1'(k) &=& \int \frac{d^3p}{(2\pi)^3 \omega_{\bf p}} \tilde{R}_1(k- p), \\
G(q|x) &=& \int \frac{d^3p d^3p'}{(2\pi)^{6} \omega_{\bf p}\omega_{\bf p'}} \tilde{R}_1\left(\frac{p+p'}{2} - q\right)  \tilde{R}_2\left(\frac{p+p'}{2} \right) e^{-i (p - p')(x_2 - x_1)}.
\eey

\section{Detector models}
In this section, we describe two simple detector models, which provide explicit expressions for the localization and reduction operators. For both models, we assume interaction terms of the for 
\bey
\hat{V} = \int d^4 x \hat{C}(x) \hat{J}(x),
\eey
that is, scalar composite operators $\hat{C}$ for the field and scalar ``current" operators for the detector. We also assume that the only observable is the spacetime point of the detection event.

\subsection{Detection through particle-like excitations}
In this model, we assume that the excited states of the detector are particle-like, in the sense that they carry momentum. Then, $\hat{J}(x)$ can be identified with an effective scalar field for these excitations,
and $R(x)$ is identified with the Wightman function
\bey
R(x) = \langle \Omega|\hat{J}(0)(\hat{I} - |\Omega\rangle \langle \Omega)\hat{J}(x)|\Omega\rangle = \langle \Omega|\hat{J}(0) \hat{J}(x)|\Omega\rangle,
\eey
assuming that $\langle \Omega|\hat{J}(0)|\Omega\rangle = 0$. 

Then, we can write $R(x)$ in the K\"ahlen-Lehmann representation \cite{Ka52, Le54}
\bey
R(x) = \int d\mu^2 \rho(\mu^2) \Delta_{(\mu)}(-x),
\eey
where $\rho(\mu^2)$ is the field's spectral density, and  
\bey
\Delta_{(\mu)}(x) = \int \frac{d^3k}{(2\pi)^3} \frac{e^{i{\bf k}\cdot {\bf x} - i \sqrt{{\bf k}^2 + \mu^2}t}}{2\sqrt{{\bf k}^2 + \mu^2} }
\eey
is the Wightman function for a free scalar field of mass $\mu$. 

We straightforwardly compute the Fourier transform
\bey
\tilde{R}({\bf k}, \omega) = \rho(\omega^2 - {\bf k}^2).
\eey
For the detection of particles with mass $m$ through absorption, 
 the absorption coefficient is $\alpha(k) = \rho(m^2)/(2k)$, and the localization operator is
 \bey
 L(k, k') = \frac{\rho\left(\frac{1}{2}(m^2 + \omega_k \omega_{k'}-kk') \right)}{\rho(m^2)}
 \eey
In the ultra-relativistic limit ($m\rightarrow 0$), $L(k, k') \rightarrow 1$, i.e., we have  maximum-localization measurements.

This model does not work for scattering-based measurements. It is straightforward to show that in this case, the diagonal matrix elements of the reduction operator, $\langle k, p|\hat{S}|k, p\rangle$, are proportional to $\rho(\mu^2)$ for negative $\mu^2$, which is zero.  

\subsection{Pointlike detectors}
In a point-like detector the dependence of $\tilde{R}(k, \omega)$ on $k$ is negligible. In this case, we can model the current operator as $\hat{J}(x) = \delta^3({\bf x})\hat{J}(t)$. Then, it is straightforward to show that
\bey
\tilde{R}(\omega) = \sum_r |\langle \Omega|\hat{J}(0)|E, r\rangle|^2
\eey
where $|E, r\rangle$ stands for the excited energy eigenstates of the detector, with $r$ the degeneracy parameter. The formulas for the localization operator and the reduction operator remain the same, with  $\tilde{R}(\omega)$ in place of $\tilde{R}(k, \omega)$. 

In Refs. \cite{QTP1, QTP3}, it was shown that typically $R(t)$ must decay with a characteristic time scale $\tau$, so that histories with respect to the detection time decohere, and the probabilities for measurement outcomes are well defined. We can consider a Lorentzian ansatz
\bey
R(t) = \frac{B}{\pi}\frac{\tau}{t^2 + \tau^2},
\eey
for some constant $B$. Then, $\tilde{R}(\omega) = Be^{-|\omega| \tau}$. This means that for detection through absorption, $L(k, k') = 1$, that is, we have maximum localization.

For detection through scattering, we find that
\bey
\tilde{R}'(\omega) = \int_0^{\sqrt{\omega^2- m^2}} \frac{dp}{\omega_p} \tilde{R}(\omega - \omega_p) =  B N_{m \tau}\left(\frac{\omega}{m}\right),
\eey
where 
\bey
N_{\alpha}(x) = \int_0^{x-1} \frac{dy}{\sqrt{(x-y)^2-1}} e^{-\alpha y}.
\eey
In the physically relevant regime, $\alpha >> 1$, in which case, the integral above is dominated by values of $y$ near zero. Therefore, we can approximate $\sqrt{(x-y)^2-1} $ with $\sqrt{x^2-1}$, and set the upper limit of integration to infinity. We obtain $N_{\alpha}(x) \simeq (\alpha \sqrt{x^2-1})^{-1}$. Then,
\bey
\tilde{R}'(\omega_k) = \frac{B}{\tau k}.
\eey
This means that the associated localization operator is 
\bey
L(k, k') = \frac{k+k'}{2\sqrt{kk'}}, 
\eey
and the reduction operator is
\bey
\langle k, q|\hat{S}|k',q'\rangle = \tau \sqrt{\frac{kk'}{\omega_q \omega_{q'}}}e^{-\frac{1}{2}\tau (\omega_k + \omega_{k'} - \omega_p - \omega_{p'})} \theta(\omega_k + \omega_{k'} - \omega_p + \omega_{p'}).
\eey
The diagonal elements are
\bey
\langle k, q|\hat{S}|k,q\rangle = \frac{\tau k}{\omega_q} e^{-\tau(\omega_k - \omega_{q})} \theta(k - q) .
\eey
We note that for $\tau \rightarrow \infty$, scattering becomes elastic.

\end{appendix}

\end{document}